\documentclass[aps,prd,showpacs,nobibnotes,nofootinbib,twocolumn,preprintnumbers,nobalancelastpage,superscriptaddress]{revtex4}
\usepackage{graphicx}
\usepackage{latexsym}
\usepackage{dcolumn}
\usepackage{bm}
\input{epsf}
\newcommand{\be}{\begin{equation}}
\newcommand{\ee}{\end{equation}}
\newcommand{\ba}{\begin{eqnarray}}
\newcommand{\ea}{\end{eqnarray}}

\newcommand{\bfi}{\begin{figure}
\epsfxsize=9cm
\epsffile}
\newcommand{\efi}{\end{figure}}
\newcommand{\bi}{\begin{itemize}}
\newcommand{\ei}{\end{itemize}}

\newcommand{\app}{Astropart. Phys. }

\newcommand{\prevd}{Physics Rev. D }

\newcommand{\nima}{Nucl. Instrum. Methods Phys. Res., Sect. A }
\newcommand{\etal}{\emph{et al.}}
\usepackage{epsfig}
\begin{document}
%
\title{Observation of the Cosmic Ray Moon shadowing effect with the ARGO-YBJ experiment}

\author{B.~Bartoli}
 \affiliation{Dipartimento di Fisica dell'Universit\`a di Napoli
                  ``Federico II'', Complesso Universitario di Monte 
                  Sant'Angelo, via Cinthia, 80126 Napoli, Italy.}
 \affiliation{Istituto Nazionale di Fisica Nucleare, Sezione di
                  Napoli, Complesso Universitario di Monte
                  Sant'Angelo, via Cinthia, 80126 Napoli, Italy.}
\author{P.~Bernardini}
 \affiliation{Dipartimento di Fisica dell'Universit\`a del Salento,
                  via per Arnesano, 73100 Lecce, Italy.}
 \affiliation{Istituto Nazionale di Fisica Nucleare, Sezione di
                  Lecce, via per Arnesano, 73100 Lecce, Italy.}
\author{X.J.~Bi}
 \affiliation{Key Laboratory of Particle Astrophysics, Institute 
                  of High Energy Physics, Chinese Academy of Sciences,
                  P.O. Box 918, 100049 Beijing, P.R. China.}
\author{C.~Bleve}
 \affiliation{Dipartimento di Fisica dell'Universit\`a del Salento,
                  via per Arnesano, 73100 Lecce, Italy.}
 \affiliation{Istituto Nazionale di Fisica Nucleare, Sezione di
                  Lecce, via per Arnesano, 73100 Lecce, Italy.}
\author{I.~Bolognino}
 \affiliation{Dipartimento di Fisica Nucleare e Teorica 
                  dell'Universit\`a di Pavia, via Bassi 6,
                  27100 Pavia, Italy.}
 \affiliation{Istituto Nazionale di Fisica Nucleare, Sezione di Pavia, 
                  via Bassi 6, 27100 Pavia, Italy.}
\author{P.~Branchini}
  \affiliation{Istituto Nazionale di Fisica Nucleare, Sezione di
                  Roma Tre, via della Vasca Navale 84, 00146 Roma, Italy.}
\author{A.~Budano}
 \affiliation{Istituto Nazionale di Fisica Nucleare, Sezione di
                  Roma Tre, via della Vasca Navale 84, 00146 Roma, Italy.}
\author{A.K.~Calabrese Melcarne}
 \affiliation{Istituto Nazionale di Fisica Nucleare - CNAF, Viale 
                  Berti-Pichat 6/2, 40127 Bologna, Italy.}
\author{P.~Camarri}
 \affiliation{Dipartimento di Fisica dell'Universit\`a di Roma ``Tor 									   Vergata'', via della Ricerca Scientifica 1, 00133 Roma, Italy.}
 \affiliation{Istituto Nazionale di Fisica Nucleare, Sezione di
                   Roma Tor Vergata, via della Ricerca Scientifica 1, 
                   00133 Roma, Italy.}
\author{Z.~Cao}
 \affiliation{Key Laboratory of Particle Astrophysics, Institute 
                  of High Energy Physics, Chinese Academy of Sciences,
                  P.O. Box 918, 100049 Beijing, P.R. China.}
% ...

\collaboration{ARGO-YBJ Collaboration}

\begin{abstract}

Cosmic rays are hampered by the Moon and a deficit in its
direction is expected (the so-called \emph{Moon shadow}). The Moon
shadow is an important tool to determine the performance of an
air shower array. Indeed, the westward displacement of the shadow
center, due to the bending effect of the geomagnetic field on the propagation of cosmic rays, allows the setting of the absolute rigidity scale of the primary particles inducing the showers recorded by the detector. In addition, the shape of the shadow permits to determine the detector point spread function, while the position of the deficit at high energies allows the evaluation of its absolute pointing accuracy.

In this paper we present the observation of the cosmic
ray Moon shadowing effect carried out by the ARGO-YBJ experiment in
the multi-TeV energy region with high statistical significance (55 standard deviations).
By means of an accurate Monte Carlo simulation of the cosmic rays propagation in the Earth-Moon system, we have studied separately the effect of the geomagnetic field and of the detector point spread function on the observed shadow.
The angular resolution as a function of the particle multiplicity and the pointing accuracy have been obtained.
The primary energy of detected showers has been estimated by measuring the westward displacement as a function of the particle multiplicity, thus calibrating the relation between shower size and cosmic ray energy.
The stability of the detector on a monthly basis has been checked by monitoring the position and the deficit of the Moon shadow.
Finally, we have studied with high statistical accuracy the shadowing effect in the "day"/"night" time looking for possible effect induced by the solar wind. 
\end{abstract}

\maketitle

\section{Introduction}

Cosmic rays (CRs) blocked in their way to the Earth by the Moon generate a deficit in its direction usually mentioned as \textit{`Moon shadow'}. The analysis of the Moon shadow observed by an air shower array may provide unique information on its performance.
At high energies, the Moon shadow would be observed by an ideal
detector as a 0.52$^{\circ}$ wide circular deficit of events,
centered on the Moon position\footnote{Actually, the width of the Moon disc ranges from 0.50$^{\circ}$ to 0.58$^{\circ}$ depending on its distance from the Earth.}. 
The actual shape of the deficit as reconstructed by the detector allows the determination of the angular resolution while the position of the deficit allows the evaluation of the absolute pointing accuracy.
In addition, charged particles are deflected by the geo magnetic field (GMF) by an angle depending on the energy. As a consequence, the observation of the displacement of the Moon shadow at low rigidities can be used to determine the relation between the shower size and the primary energy.

The same shadowing effect can be observed in the direction of the
Sun but the interpretation of this phenomenology is less straightforward.
In fact, the displacement of the shadow from the apparent
position of the Sun could be explained by the joint effects of the
GMF and of the solar and interplanetary magnetic fields (SMF and IMF, respectively), whose configuration considerably changes with the phases of the solar activity cycle \cite{amenomori-sun00}.
In this regard, understanding the Moon shadow phenomenology is a useful tool to investigate the GMF features needed to disentangle the effects of different magnetic fields on the Sun shadow and to perform a measurement of the IMF during a minimum of the solar activity \cite{argo-solar}. 

Finally, the Moon shadow can be exploited to measure the antiproton content in the primary CRs. In fact, acting the Earth-Moon system as a magnetic spectrometer, paths of primary antiprotons are deflected in the opposite direction with respect to those of the protons in their way to the Earth. This effect has been used to set limits on the antiproton flux at TeV energies not yet accessible to balloon or satellite experiments \cite{macro03,l3,tibet07,antip}.

In this paper we present the observation of the cosmic ray Moon shadowing effect carried out by the ARGO-YBJ experiment during the period from July 2006 to November 2010.
We report on the angular resolution, the pointing accuracy and the rigidity scale calibration of the detector in the multi-TeV energy region. The results are compared with the predictions of a detailed simulation of cosmic ray propagation in the Earth-Moon system.

The paper is organized as follows. In Sec. II the ARGO-YBJ detector is described and the event reconstruction sketched out. In Sec. III the data analysis performed with two different background estimation techniques is outlined. The results of a Monte Carlo simulation of the cosmic ray propagation in the Earth-Moon system are presented in Sec. IV. The measurement of the pointing accuracy and of the angular resolution as well as the evaluation of the absolute rigidity scale are discussed in Sec. V. A high statistics study of the day-night effect is also reported in Sec. V. 
A summary of the obtained results is given in Sec. VI.

\section{The ARGO-YBJ experiment}

\subsection{The detector}

The ARGO-YBJ experiment, located at the YangBaJing Cosmic Ray
Laboratory (Tibet, P.R. China, 4300 m a.s.l., 606 g/cm$^2$), is currently the only air shower array exploiting the full coverage approach at very high
altitude, with the aim of studying the cosmic radiation at an
energy threshold of a few hundred GeV.

The detector is constituted by a central carpet $\sim$74$\times$
78 m$^2$, made of a single layer of resistive plate chambers
(RPCs) with $\sim$93$\%$ of active area, enclosed by a guard ring
partially instrumented ($\sim$20$\%$) up to $\sim$100$\times$110
m$^2$. The apparatus has a modular structure, the basic data
acquisition element being a cluster (5.7$\times$7.6 m$^2$),
made of 12 RPCs (2.85$\times$1.23 m$^2$ each). Each chamber is
read by 80 external strips of 6.75$\times$61.80 cm$^2$ (the spatial pixels),
logically organized in 10 independent pads of 55.6$\times$61.8
cm$^2$ which represent the time pixels of the detector \cite{aielli06}. 
The readout of 18360 pads and 146880 strips is the experimental output of the detector. The relation between strip and pad multiplicity has been measured and found in fine agreement with the Monte Carlo prediction \cite{aielli06}.
In addition, in order to extend the dynamical range up to PeV energies, each chamber is equipped with two large size pads (139$\times$123 cm$^2$) to collect the total charge developed by the particles hitting the detector \cite{bigpad}.
The RPCs are operated in streamer mode by using a gas mixture (Ar 15\%, Isobutane 10\%, TetraFluoroEthane 75\%) for high altitude operation \cite{bacci00}. The high voltage settled at 7.2 kV ensures an overall efficiency of about 96\% \cite{aielli09a}.
The central carpet contains 130 clusters (hereafter ARGO-130) and the
full detector is composed of 153 clusters for a total active
surface of $\sim$6700 m$^2$ (Fig. \ref{fig:argo-layout}). The total instrumented area is $\sim$11000 m$^2$.
%
\bfi{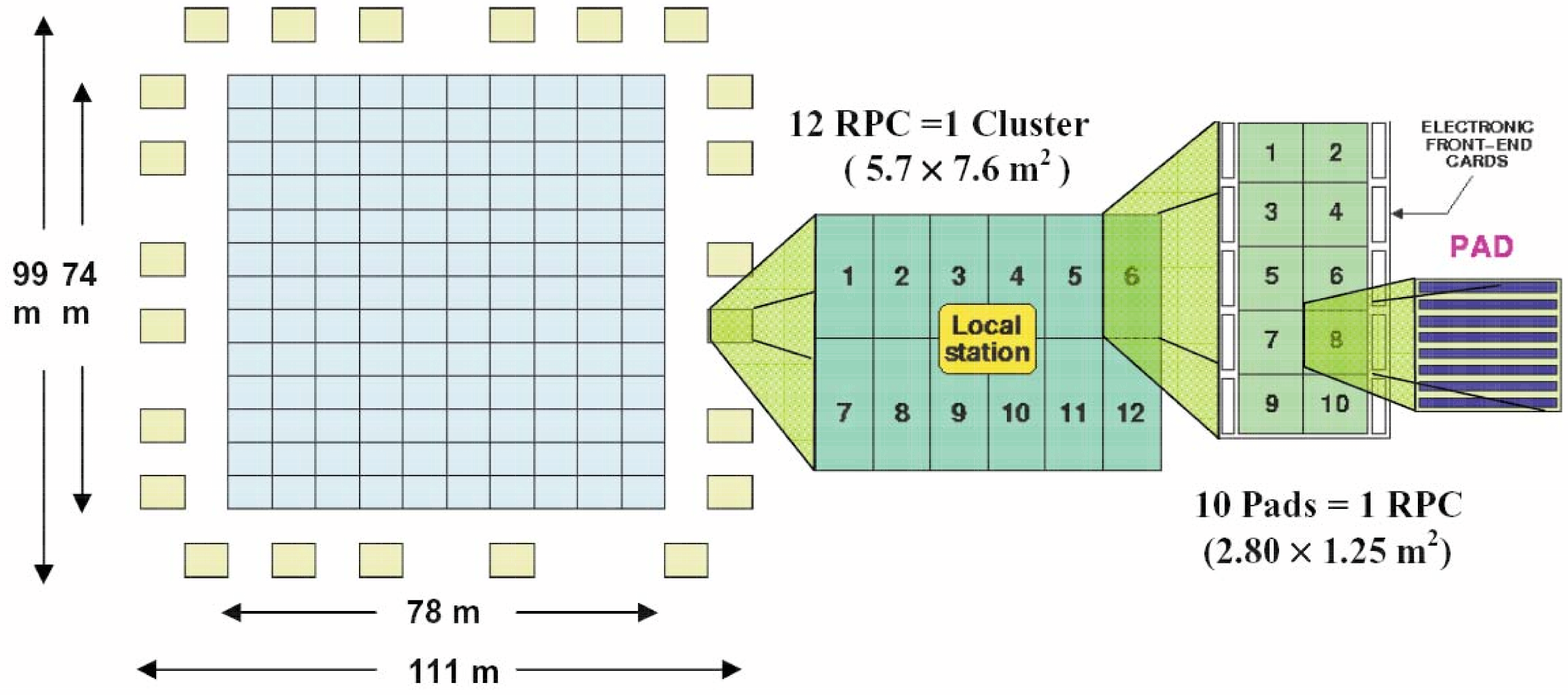}\\
    \caption{Layout of the ARGO-YBJ experiment (see text for a detailed description of the detector).
    \label{fig:argo-layout}}
\efi
%
The information on strip multiplicity and the arrival times recorded by each pad are received by a local station devoted to manage the
data of each cluster. A central station collects the information of
all the local stations. 
The time of each fired pad in a window of 2 $\mu$s around the trigger time
and its location are used to reconstruct the position of the shower core and the arrival direction of the primary particle.
In order to perform the time calibration of the 18360 pads, a
software method has been developed \cite{aielli09b}. To check the stability of the apparatus a control system (DCS) monitors continuously the current of each
RPC, the gas mixture composition, the high voltage distribution as well as the environment conditions (temperature, atmospheric pressure, humidity).
A simple, yet powerful, electronic logic has been implemented to build an inclusive trigger. This logic is based on a time correlation between the pad signals depending on their relative distance. In this way, all the shower events giving a number of fired pads N$_{pad}\ge$ N$_{trig}$ in the central carpet in a time window of 420 ns generate the trigger.
This trigger can work with high efficiency down to N$_{trig}$ = 20,
keeping negligible the rate of random coincidences.

Because of the small pixel size, the detector is able to record events with a particle density exceeding 0.003 particles m$^{-2}$, keeping good linearity up to a core density of about 15 particles m$^{-2}$.
This high granularity allows a complete and detailed three-dimensional reconstruction of the front of air showers at an energy threshold of a few hundred GeV. Showers induced by high energy primaries ($>$ 100 TeV) are also imaged by the analog readout of the large size pads \cite{bigpad}.

The whole system, in smooth data taking since July 2006 with ARGO-130, is in stable data taking with the final configuration of 153 clusters since November 2007 with the trigger condition N$_{trig}$ = 20 and a duty cycle $\geq$85\%. The trigger rate is $\sim$3.5 kHz with a dead time of 4$\%$.

In the present study the data recorded by the digital readout have been analyzed to measure the Moon shadow effect induced by low-energy primaries.

\subsection{Event reconstruction and data selection}

The reconstruction of the shower parameters is carried out through the
following steps.

At first, a plane surface is analytically fitted (with weights equal to 1) to the shower front. This procedure is repeated up to 5 times, each iteration rejecting hits whose arrival time is farther than 2 standard deviations from the mean of the distribution of the time residuals from the fitted plane surface. This iterative procedure is able to reject definitively from the reconstruction the time values belonging to the non-Gaussian tails of the arrival time distributions \cite{icrc05_risang}. 
After this first step the problem is reduced to the nearly-vertical case by means of a projection which makes the fit plane overlapping the detector plane. 
Thereafter, the core position, i.e. the point where the shower axis intersects the detection plane, is obtained fitting the lateral density distribution of the secondary particles to a modified Nishimura-Kamata-Greisen (NKG) function. The fit procedure is carried out via the maximum likelihood method \cite{llf}.
Finally, the core position is assumed to be the apex of a conical surface to be fitted to the shower front. The slope of such a conical correction is fixed to $\alpha$ = 0.03 ns/m \cite{icrc05_risang}.

The capability of reconstructing the primary arrival direction 
can be further enhanced by applying robust statistical methods 
in the analysis of the shower front, conveniently weighting the 
contribution of the most delayed particles. In detail, we first fit a conical surface to the shower image, by minimizing the sum of the squares of the time residuals. At this stage, all the particles hitting the detector have the same weight $w_i$=1. After computing the RMS of the time residual distribution with respect to such a conical surface, we set K = 2.5$\cdot$RMS as a `scale parameter' and perform the minimization of the square of the time residuals weighted sum, where $w_i$=1 if the particle is onward the shower front, $w_i$=$f((t_i^{exp}-t_i^{fit})/K)$ otherwise. The function $f(x)$ is a common Tukey biweight function \cite{Tukey}. The fit procedure is iterated, every time refreshing the scale parameter, until the last reconstructed direction differs from the previous one for less than 0.1$^\circ$. 

The analysis reported in this paper refers to events selected according to
the following criteria: (1) more than 20 strips N$_{strip}$ should be fired on the ARGO-130 carpet; (2) the zenith angle of the shower arrival direction should be less than 50$^{\circ}$; (3) the reconstructed core position should be inside an area 150$\times$150 m$^2$ centered on the detector. 
After these selections the number of events analyzed is about 2.5$\times$10$^{11}$ (about 10$^9$ inside a 10$^{\circ}\times$10$^{\circ}$ angular region centered on the Moon position).
According to simulations, the median energy of the selected protons is E$_{50}\approx$1.8 TeV (mode energy $\approx$0.7 TeV).

\section{Moon shadow analysis}

For the analysis of the shadowing effect three different sky maps in celestial coordinates (right ascension R.A. and declination DEC.) are built: the \emph{event} and \emph{background} maps with 0.1$^{\circ}\times$0.1$^{\circ}$ bin size, necessary to determine the deficit shape, and the \emph{significance} map used to estimate the statistical significance of the observation.

The event map, of size 10$^{\circ}\times$10$^{\circ}$, centered on the Moon location, is filled with the detected events. 
Cosmic rays blocked by the Moon have to be as many as the background events lying within a region as large as the Moon disc. 
A suitable background estimation is therefore a crucial point of the analysis.
The background has been evaluated with both the \emph{time-swapping}
\cite{alexandreas} and the \emph{equi-zenith angle} \cite{amenomori93} methods in order to investigate possible systematic uncertainties in the background calculation.

In the time-swapping method, N \emph{"fake"} events are generated for each detected one, every time replacing the measured arrival time with a new one. Such a random time spans over a 3 hours wide buffer of recorded data, to minimize the systematic effects induced by environment variations (i.e. temperature and atmospheric pressure). Changing the time, each fake event maintains the same declination of the original one, but has a different right ascension. In this way, a new sky map (the background map) is built.
If the number of fake events generated for each event is N, the fluctuations of the background estimation are reduced of a factor $\sim\sqrt{N}$ with respect to those of the signal. In this analysis we set N = 10.
The strong point of the time-swapping technique is that it takes into account only the sky region where the Moon actually passes through, though a few tens of minutes before or later. On the other hand, also the time of all the events is swapped, then the obtained background at the Moon position is slightly underestimated and thus the signal is underestimated. This underestimate ranges from $\sim$4\% to 10$\%$, increasing with the angular resolution, hence depending on the event multiplicity. The observed event rate is then corrected using the appropriate factor \cite{aielli10}.

With the equi-zenith angle method the number of cosmic rays recorded in the off-source cells with the same size, at the same zenith angle and in the same time intervals as the on-source cell is averaged.
The method is able to eliminate various spurious effects caused by instrumental and environmental variations, such as changes in pressure and
temperature that are hard to control and tend to introduce systematic errors in the measurement.
The equi-zenith background estimation is achieved in the reference frame of the experiment, i.e. using the local coordinates zenith and azimuth. The Moon position is computed every minute and 6 off-source bins are symmetrically aligned on both sides of the on-source field, at the same zenith angle. The nearest off-source bins are set at an azimuth distance 5$^{\circ}$ from the on-source bin. The other off-source bins are located every 5$^{\circ}$ from them. The average of the event densities inside these bins is taken to be the background. The equi-zenith technique uses only showers detected at the same time of the on-source events and permits to take into account every minimal sudden environment change. Nonetheless, its efficiency relies on the assumption that the events triggering the detector are uniformly distributed in azimuth, which is true only at the first order. As a matter of fact, different factors can induce a modulation in the event distribution. The GMF, for example, induces a modulation as large as $\sim$1\% for low energy showers \cite{ivanov,hhh}, making necessary a proper correction to the background estimation.

The significance map is obtained from the event and background maps after applying the following smoothing procedure to take into account the angular resolution of the detector. 
The bins of the maps are `integrated' over a circular area of radius $\psi$, i.e. every bin is filled with the content of all the surrounding bins whose center is closer than $\psi$ from its center. The value of $\psi$ is related to the angular resolution of the detector, and corresponds to the
radius of the observational window that maximizes the signal to
background ratio, which in turn depends on the event multiplicity: when the point spread function (PSF) is a Gaussian with RMS $\sigma$, then $\psi$ = 1.58$\cdot\sigma$ and contains $\sim$72$\%$ of the events. The optimal size of the observational window as a function of the event multiplicity is obtained from the analysis of the event map and compared with the results of a Monte Carlo simulation (Sec. V, B).

After such a smoothing procedure, an integrated \emph{``source map''} is obtained by subtracting the integrated background map content from that of the integrated event map. The deficit significance of each bin of the source map with respect to the content of the corresponding background map bin is computed according to Li and Ma \cite{li-ma}, providing the \emph{``significance map''}. This map is used to estimate the statistical significance of the observation.

A detailed study of the two background calculation methods in the same sky region has shown that on average they give significances of the deficit consistent within about 1 standard deviation, corresponding to a few per cent of uncertainty on the number of events in the observed Moon shadow signal.

In the following the results obtained with the equi-zenith method are shown.

In Fig. \ref{fig:moon-wangbo} the significance map of the Moon region observed with data recorded in the period July 2006 - November 2009 (about 3200 hours on-source) is shown for events with fired strips N$_{strip}>$ 100. The opening angle $\psi$ used in the smoothing procedure is 1$^{\circ}$. The statistical significance of the maximum deficit is about 55 standard deviations. The ARGO-YBJ experiment is observing the Moon shadow with a significance of about 9 standard deviations per month.
%
\bfi{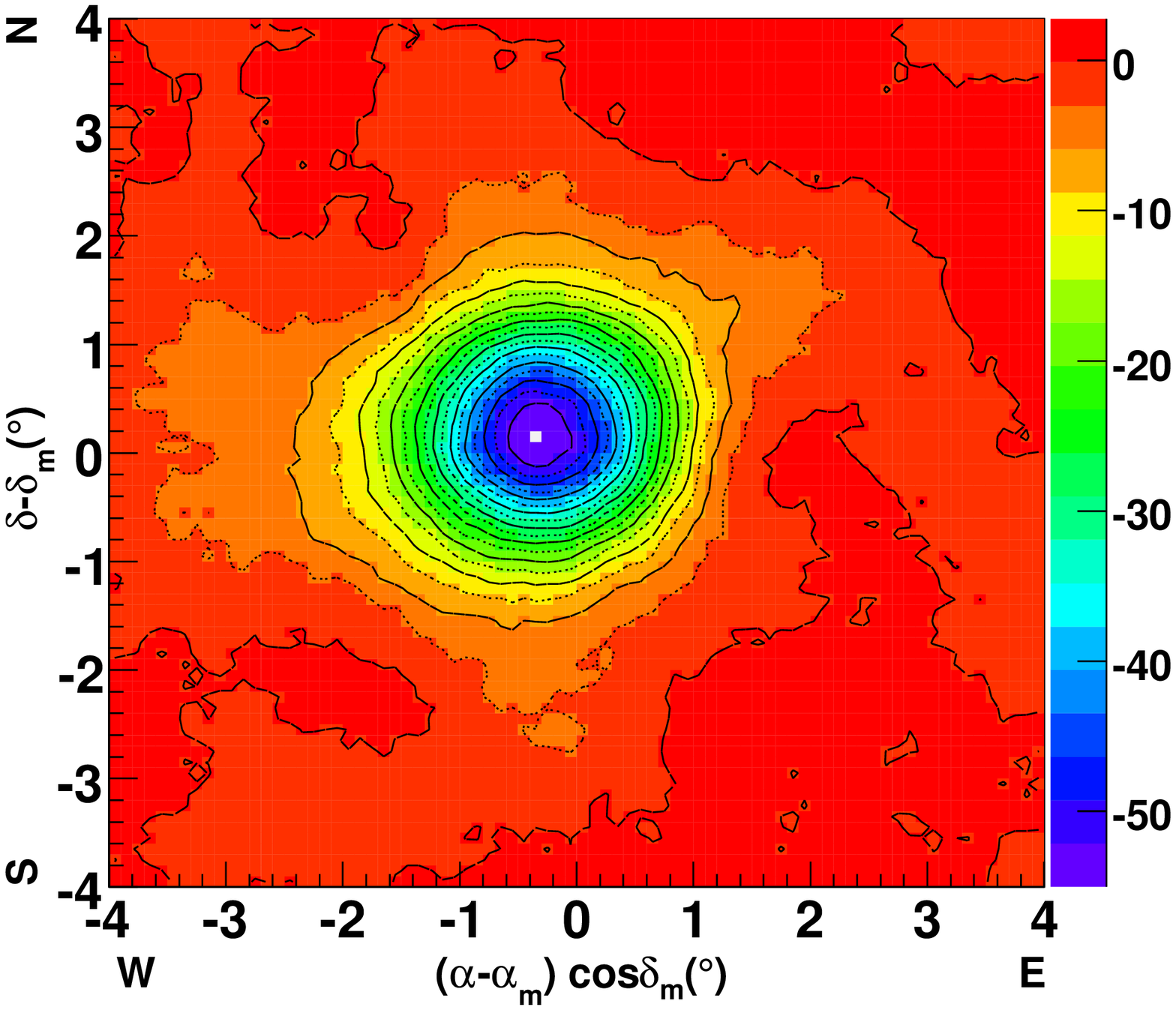}\\
 \caption{Significance map of the Moon region for events with N$_{strip}>$100, observed by the ARGO-YBJ experiment in the period July 2006 - November 2009 (about 3200 hours on-source in total). The coordinates are R.A. $\alpha$ and DEC. $\delta$ centered on the Moon position ($\alpha_m$, $\delta_m$).
The color scale gives the statistical significance in terms of standard deviations.
  \label{fig:moon-wangbo}}
\efi
%

%
\bfi{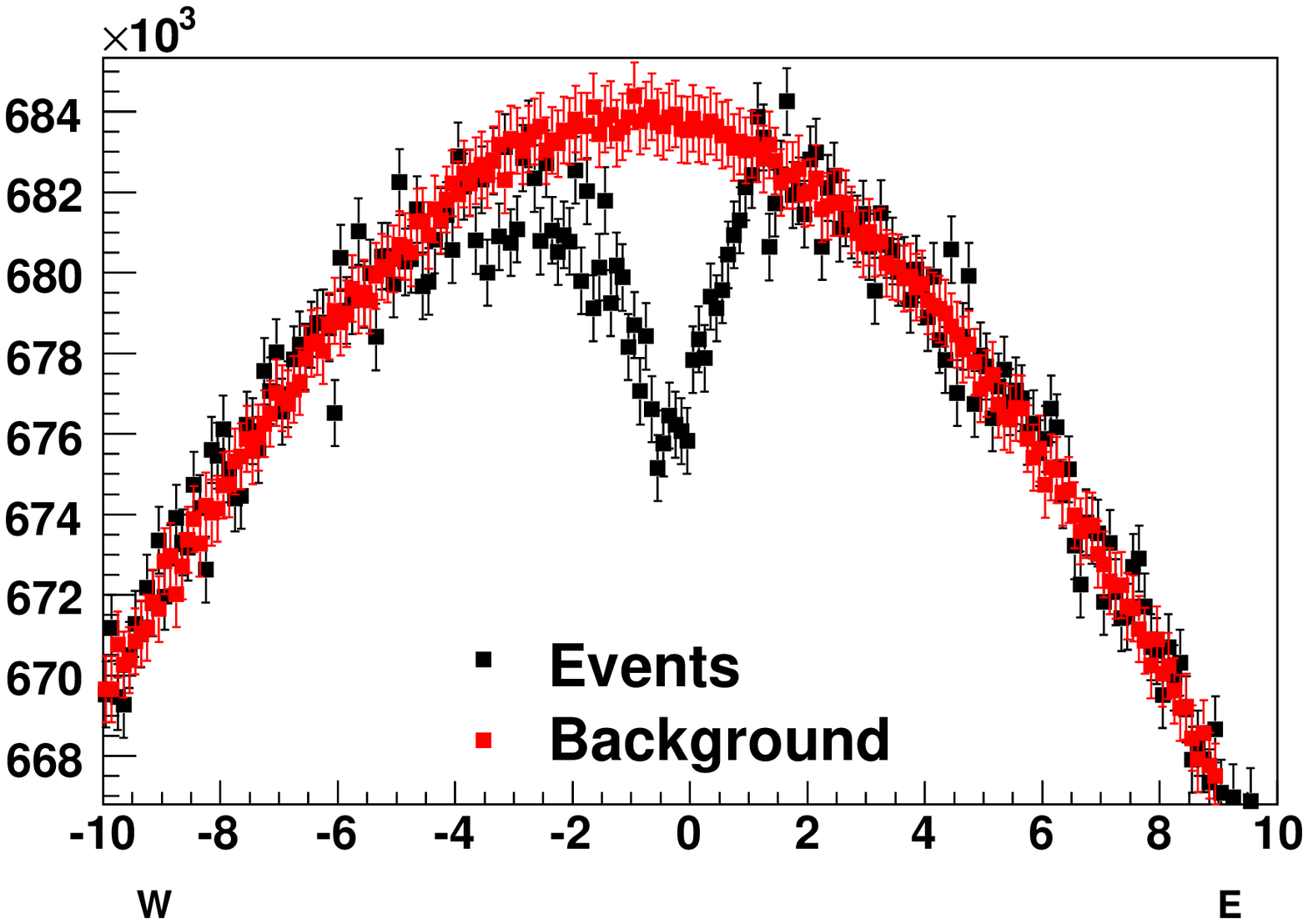}\\
 \caption{Deficit of CRs around the Moon position projected along the R.A. direction. Showers with N$_{strip}>$100 recorded in the period July 2006 - November 2009 are shown.
  \label{fig:moon-evbkg}}
\efi
%
As can be noticed from Fig. \ref{fig:moon-evbkg}, the Moon shadow turns out to be a deep in the smooth CR signal observed by ARGO-YBJ, even without subtracting the background contribution. The background events are not uniformly distributed around the Moon, because of the non uniform exposure of the map bins to CR radiation.

\section{Monte Carlo simulation}

A detailed Monte Carlo simulation has been performed in order to propagate the CRs in the Earth-Moon system \cite{mcmoon}.
The air shower development in the atmosphere has been generated
with the CORSIKA v. 6.500 code \cite{corsika}. 
The electromagnetic interactions are described by the EGS4 package while the hadronic interactions above 80 GeV are reproduced by the QGSJET-II.03 and the SYBILL models. The low-energy hadronic interactions are described by the FLUKA package. Cosmic ray spectra have been simulated in the energy range from 10 GeV to 1 PeV following the relative normalization given in \cite{horandel}. About 10$^8$ showers have been sampled in the zenith angle interval 0$^{\circ}$ - 60$^{\circ}$. The secondary particles have been propagated down to cut-off energies of 1 MeV (electromagnetic component) and 100 MeV (muons and hadrons).
The experimental conditions (trigger logic, time resolution, electronic noises, relation between strip and pad multiplicity, etc.) have been taken into account via a GEANT4-based code \cite{geant4}. The core positions have been randomly sampled in an energy-dependent area large up to 2$\cdot$10$^3$ $\times$ 2$\cdot$ 10$^3$ m$^2$, centered on the detector. Simulated events have been generated in the same format used for the experimental data and analyzed with the same reconstruction code.

\subsection{The geomagnetic model}

To properly describe the Moon shadowing effect, the magnetic field from the Moon to the Earth must be taken into account as much accurately as possible. Since the contribution due to the Moon itself is negligible, the total field turns out to be the superposition of the IMF, due to the solar wind, and the GMF. The latter is by far the most intense acting upon the particles propagating in the relatively narrow region between the Moon and the Earth. Therefore, the observed deviation of the CR trajectories depends mainly on the experimental site position relative to the GMF.

It has been already noticed that if a primary cosmic ray (energy $E$, 
charge $Z$) traversing the GMF is observed by a detector 
placed at YangBaJing, its trajectory is bent along the East-West direction, whereas no deviation is expected along the North-South one \cite{amenomori-sun00}. To a first approximation, the amount of the East-West shift can be written as:
%
\begin{equation}
\label{eq:DipoleDisplacement}
\Delta\alpha\simeq-1.58^{\circ}\frac{Z}{E[\textrm{TeV}]}
\end{equation}
%
The sign is set according to the usual way to represent the East-West
projection of the Moon maps (see Figs. \ref{fig:moon-wangbo} and \ref{fig:SimulatedMaps}). Equation (\ref{eq:DipoleDisplacement}) can be easily derived by assuming that the GMF is due to a pure static dipole lying in the center of the Earth (see Appendix).
As shown below, Eq. (\ref{eq:DipoleDisplacement}) is valid for nearly vertical primaries with energy greater than a few TeV.
%
\bfi{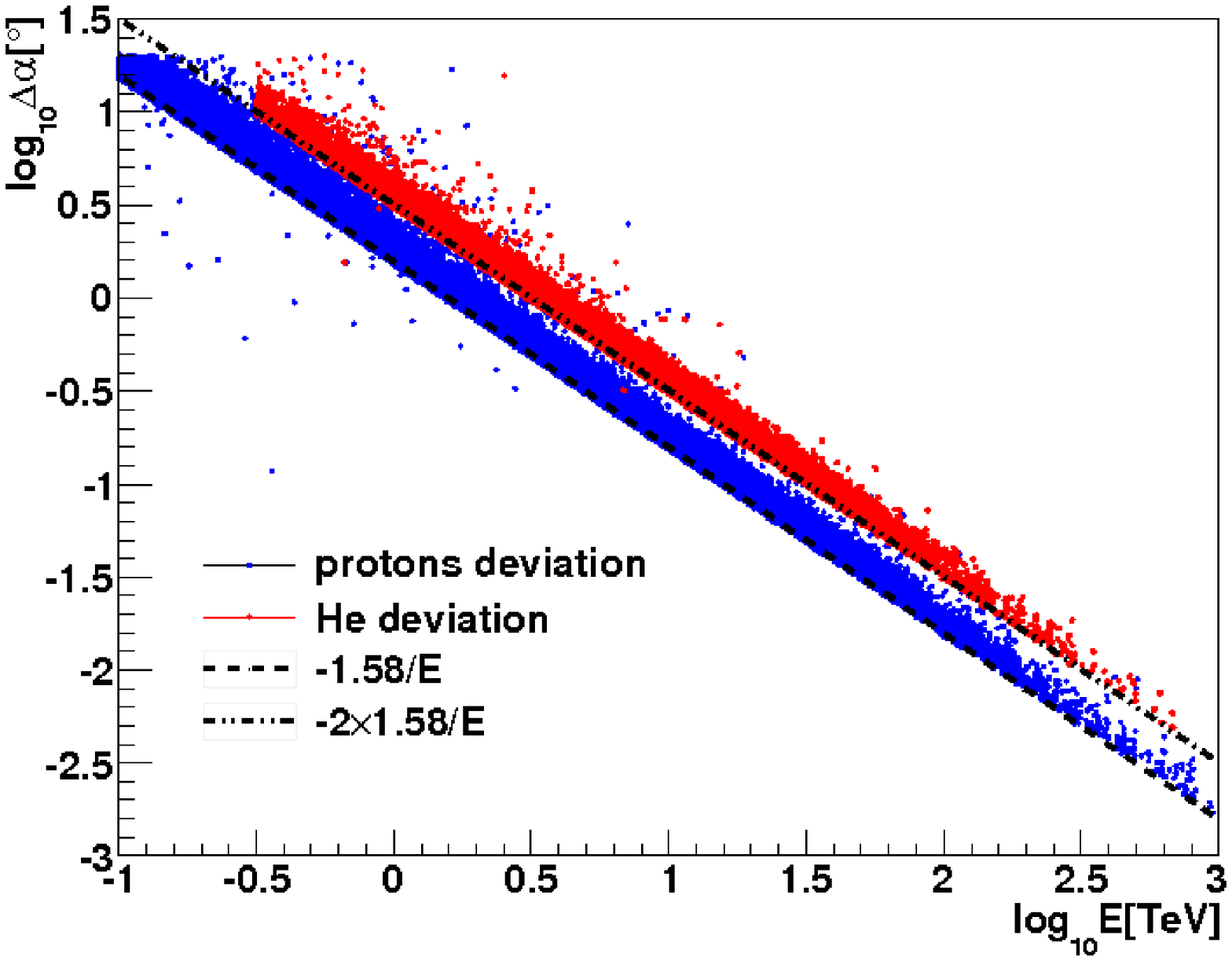}\\
 \caption{Deviation induced by the GMF on protons (blue points) and Helium nuclei (red points). Each point refers to a simulated primary. The analytical trends obtained from Eq. (\ref{eq:DipoleDisplacement}) are also shown as dashed (protons) and dot-dashed (He) lines.
   \label{fig:MagneticModels}}
\efi
%
To perform an evaluation of the bending effect, it is necessary
to adopt a model of the magnetic field in the Earth-Moon system. Such a model provides an estimation of the coefficients of the magnetic field expansion in spherical harmonics. The simplest one is the so-called virtual dipole model (VDM) \cite{VDM}. A better choice is the Tsyganenko-IGRF model (hereafter T-IGRF) \cite{Tsyganenko}, which takes into account both internal
and external magnetospheric sources by using data available from spacecraft missions. We compared the effect on the particle trajectories of VDM and T-IGRF, finding in both cases non negligible differences with respect to Eq. (\ref{eq:DipoleDisplacement}), which underestimates the deviation up to 10 - 15$\%$, mostly for low energy primaries. Among the two models themselves, we observed discrepancies up to $\sim 10\%$, corresponding to 0.4$^{\circ}$ -  0.7$^{\circ}$ for sub-TeV primary energies, mainly due to the description of the field intensity nearby the Earth surface. Since the T-IGRF model takes into account more factors, we will refer to it hereafter.

In Fig. \ref{fig:MagneticModels} the actual East-West displacement obtained applying the T-IGRF model to the propagation of both protons (Z=1) and Helium nuclei (Z=2) can be appreciated.
The points represent the deflection undergone by a nucleus propagating from the Moon to the YangBaJing geographical site according to the following simulation procedure: (1) the primary energy is sampled according to the spectra quoted in \cite{horandel}; (2) the arrival direction is sampled from an isotropic distribution; (3) the events are spread uniformly during year 2008.
In this figure the lines reproduce the deviation expected from Eq. (\ref{eq:DipoleDisplacement}) for both protons and Helium nuclei. 
The analytical approach clearly underestimates the East-West deviation, in particular for sub-TeV events.
%
\bfi{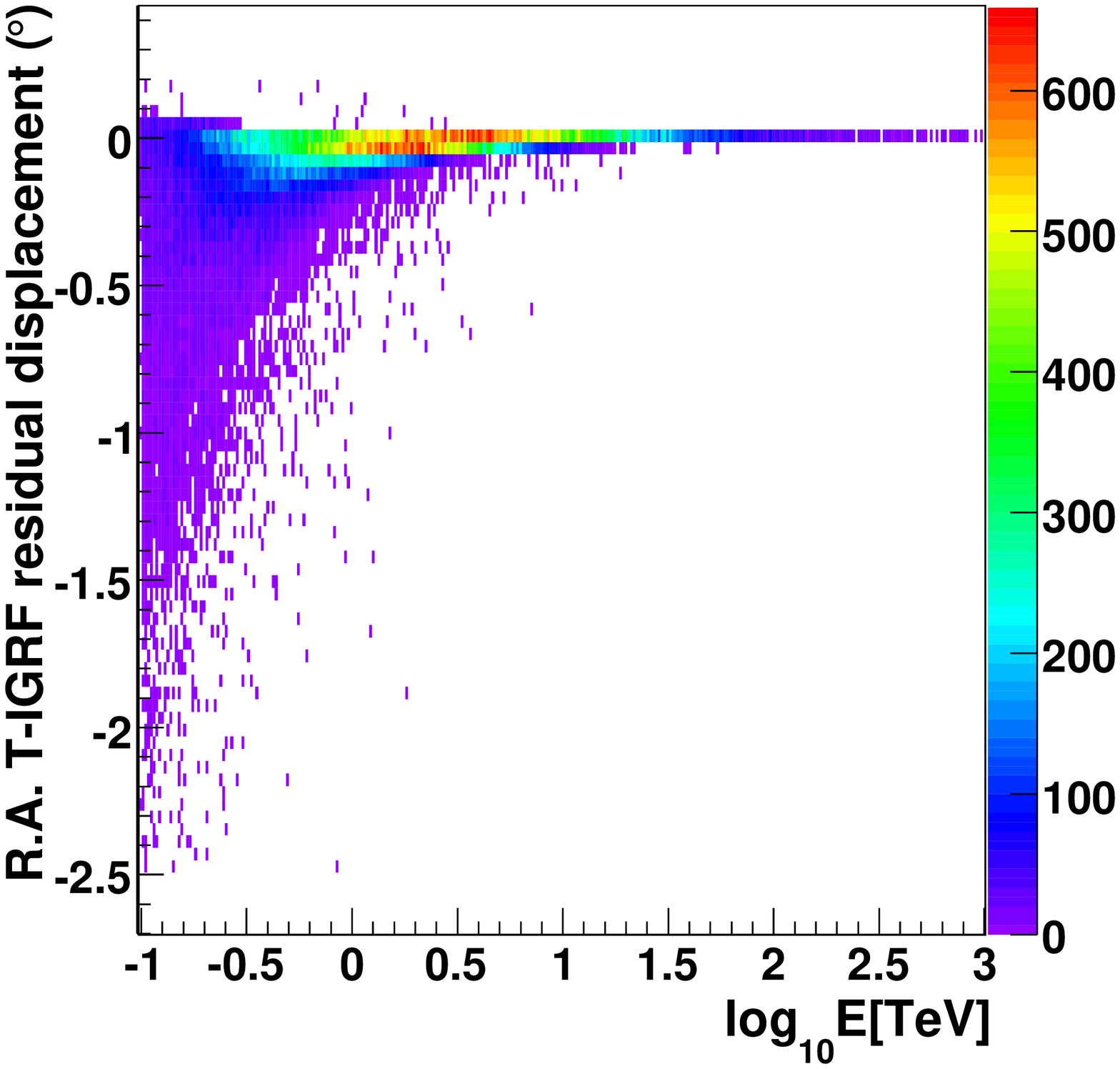}\\
(a)\\
\epsfxsize=9cm \epsffile{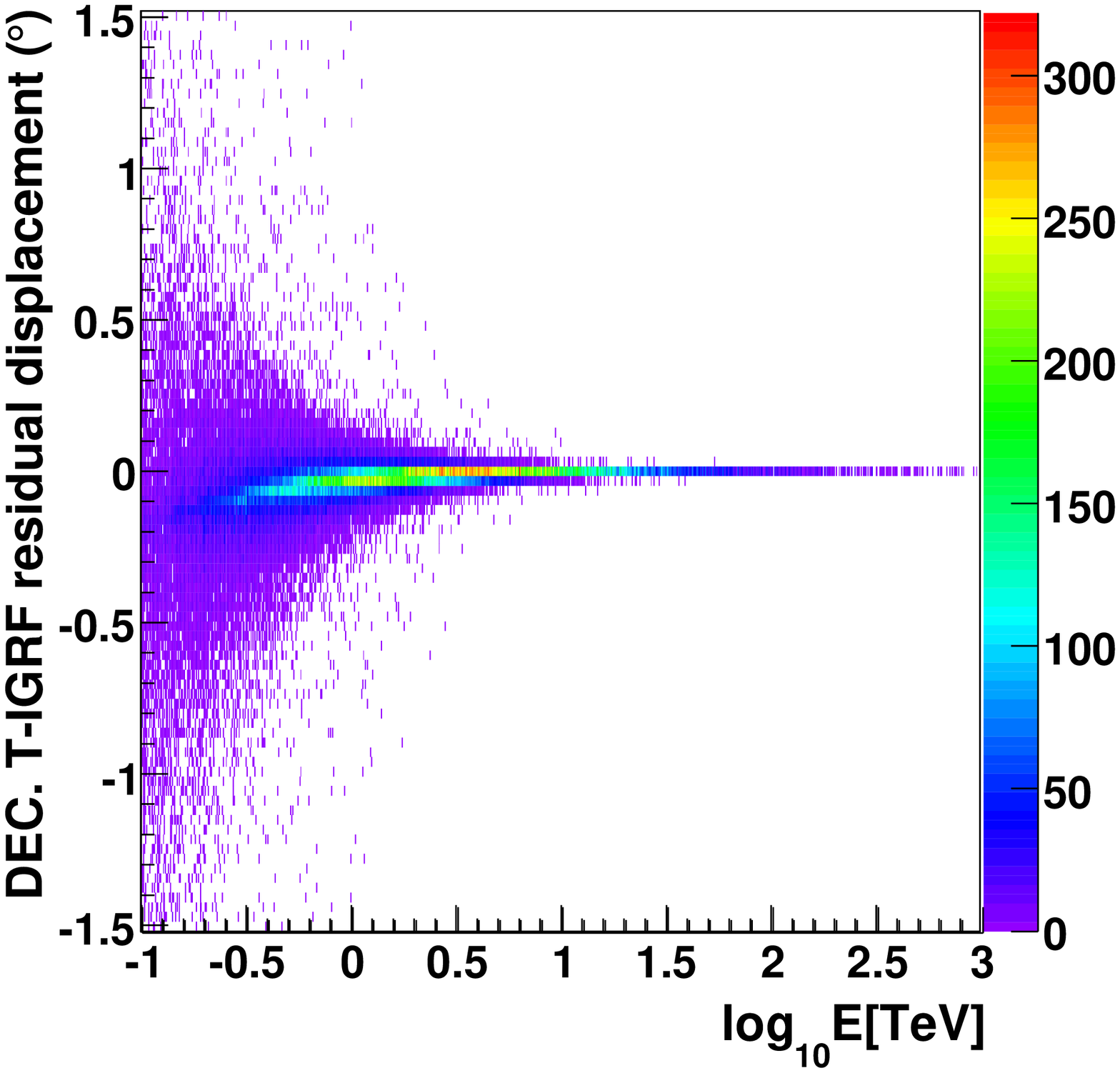}\\
(b)
\caption{Panel (a) shows the residual displacement with respect to the analytical expectation (Eq. (\ref{eq:DipoleDisplacement})) along the East-West direction as a function of the primary energy.
Panel (b) shows the residual displacement along the North-South direction.
The deviation is calculated by applying the T-IGRF model (see text). The color scale represents the number of showers lying on the single pixel of the figure.
\label{fig:GeomagneticTrend}}
\efi
%
Figure \ref{fig:GeomagneticTrend} shows the difference of the
T-IGRF-induced deviation with respect to the leading term
$1.58^{\circ}Z/E\textrm{[TeV]}$. From the plot (a) is evident that the T-IGRF model predicts a deviation along the East-West direction greater than the one expected from Eq. (\ref{eq:DipoleDisplacement}). Altough this effect is negligible for energies E$>$10 TeV, at lower energies E$<$1 TeV the difference can reach 1$^{\circ}$ or more.
The plot (b) shows the difference along the North-South direction.
Notice that unlike what the analytical approach would suggest, the North-South deviation of a primary can be non-null, being zero only on average.

\subsection{Moon shadow simulation}

By following the procedure described above, we can obtain the Moon shadow maps represented in Fig. \ref{fig:SimulatedMaps}, where the effect of folding the detector PSF and the GMF is investigated. 
After the simulation, only events satisfying the selection criteria discussed in Sec. II, B have been taken into account.

In the upper left plot the Moon disc as it would be observed by an ideal detector without any effect induced by the GMF is shown. In the upper right plot the effect of the GMF on the ideal detector is displayed.
The showers do not gather any more in the Moon disc. Along the R.A. direction (also East-West hereafter), they all suffer a `negative' deviation (what we call `westward'), inversely proportional to the energy.
The long tail of the left part of the map is due to the lowest energy CRs (sub-TeV showers) which are more deviated.
Along the DEC. direction (also North-South hereafter), a significant deviation is suffered only by the least energetic primaries, all others propagating imperturbed.
%
\bfi{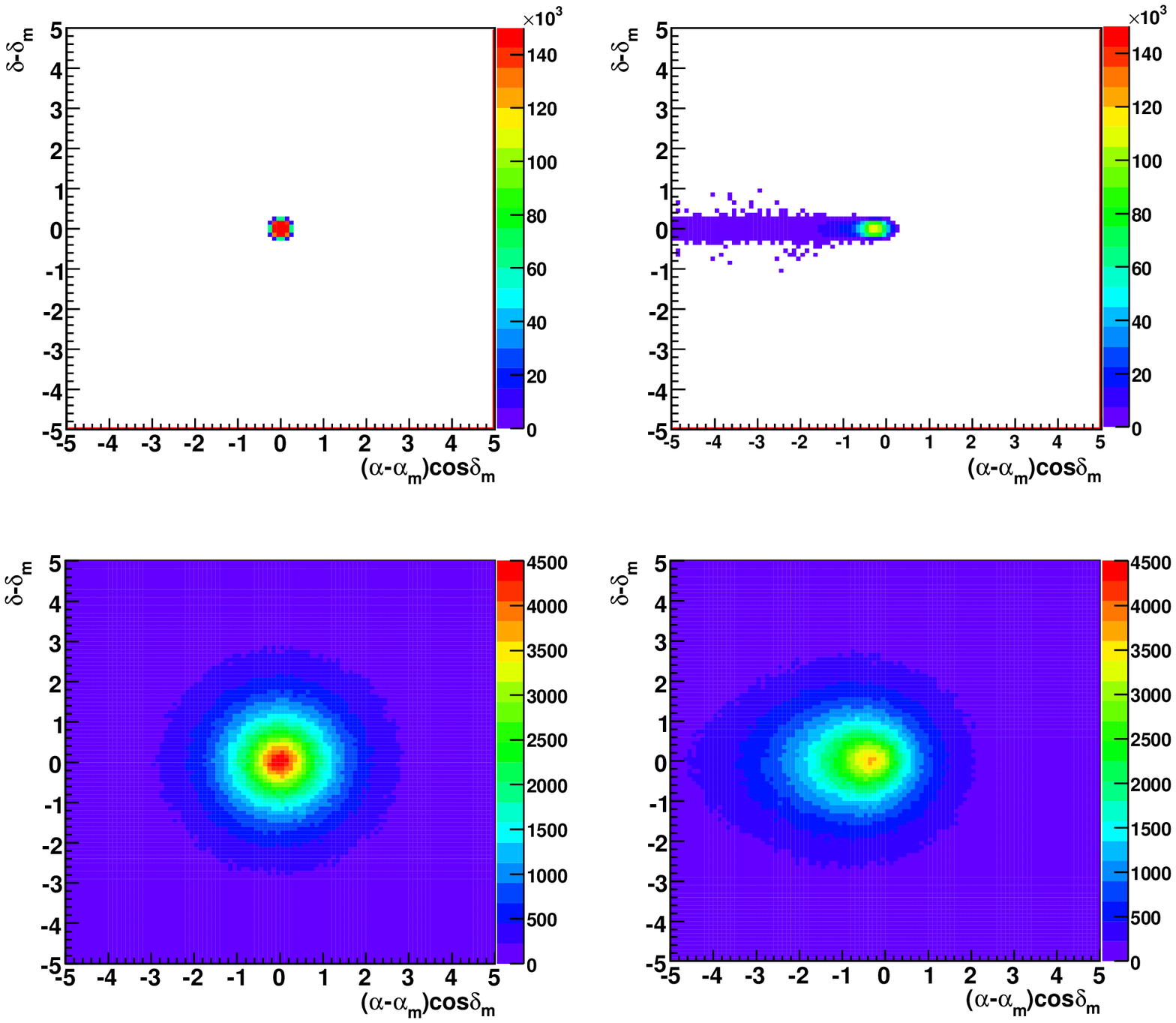}\\
  \caption{The effect of folding different contributions to the Moon signal. Upper part of the figure: Moon as it would be observed by an ideal detector without GMF (left plot). Effect of the GMF on an ideal detector (right plot). Lower part: effect of the detector PSF without and with the GMF (left and right plot, respectively). Only the showers satisfying the selection criteria in Sec. II, B are shown. The color scale represents the number of showers lying on the single pixel of the figure.
  \label{fig:SimulatedMaps}}
\efi
%
In the lower plots the effect of the detector PSF is taken into account, without and with the GMF. As it can be seen from the bottom left plot, the detector PSF only smears out the signal, leaving intact the circular symmetry, as expected. The combined effect of the GMF and the detector PSF is shown in the bottom right plot.
%
\bfi{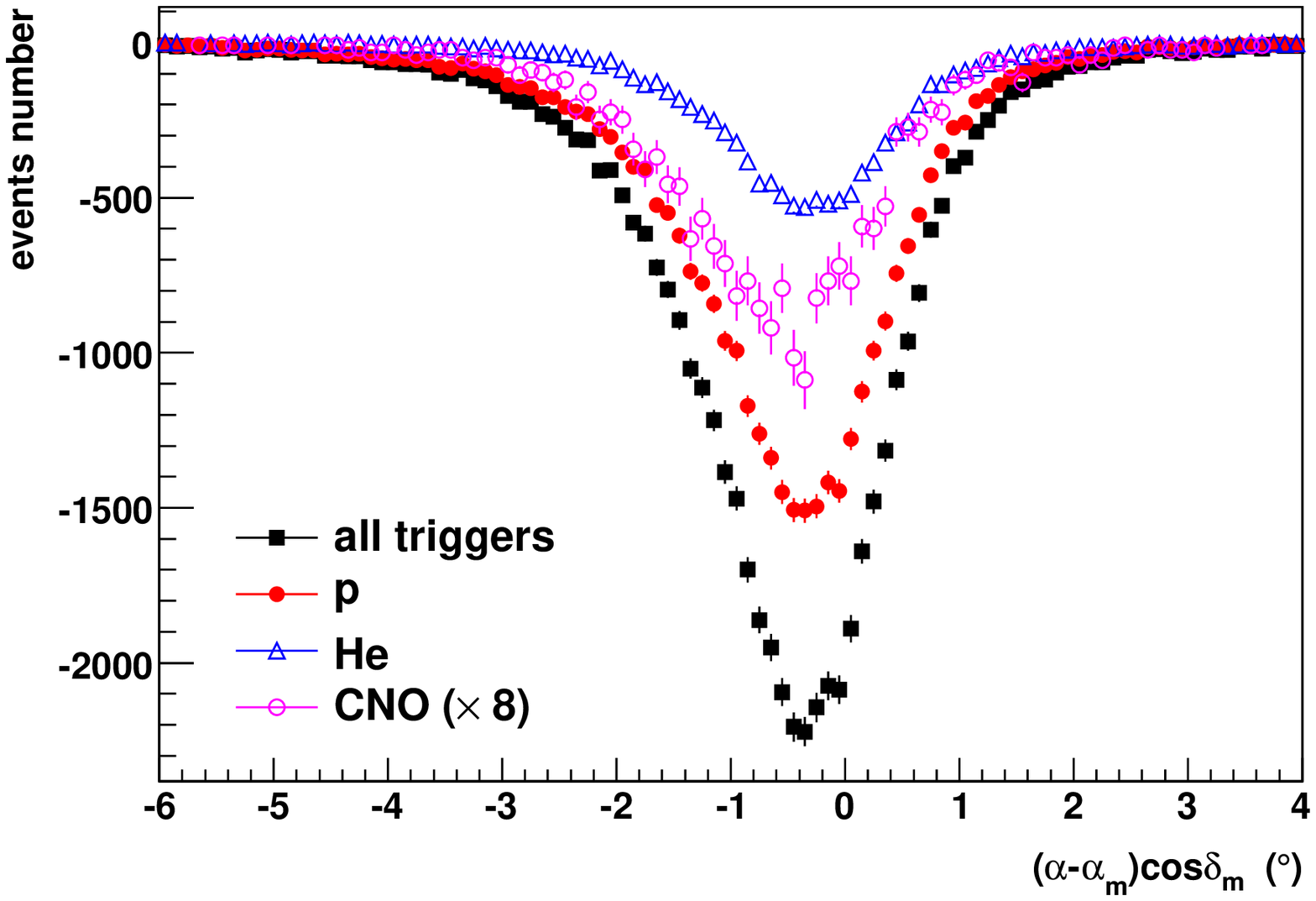}\\
  \caption{Simulated deficit counts around the Moon projected to the East-West axis for N$_{strip}>$100. The contribution of different primaries to the Moon shadow deficit can be appreciated. The CNO component has been multiplied by a factor 8.
  \label{fig:components}}
\efi
%
The contribution of different cosmic ray primaries (protons, Helium and CNO group) to the Moon shadow deficit is shown in Fig. \ref{fig:components}. 
Events contained in an angular band parallel to the East-West axis and centered on the observed Moon position, compatible with the multiplicity-dependent angular resolution, are used.
According to Fig. \ref{fig:MagneticModels}, the displacement of Helium-induced showers is expected to be greater than that of showers generated by proton primaries.
This result is not evident in Fig. \ref{fig:components}. Indeed, the analysis criteria based on the event multiplicity (the experimental observable) select the rigidity distributions shown in Fig. \ref{fig:enerigidity}. The Helium rigidity spectrum exhibits a mode higher than that of the proton rigidity spectrum, resulting in a lower displacement. 
%
\bfi{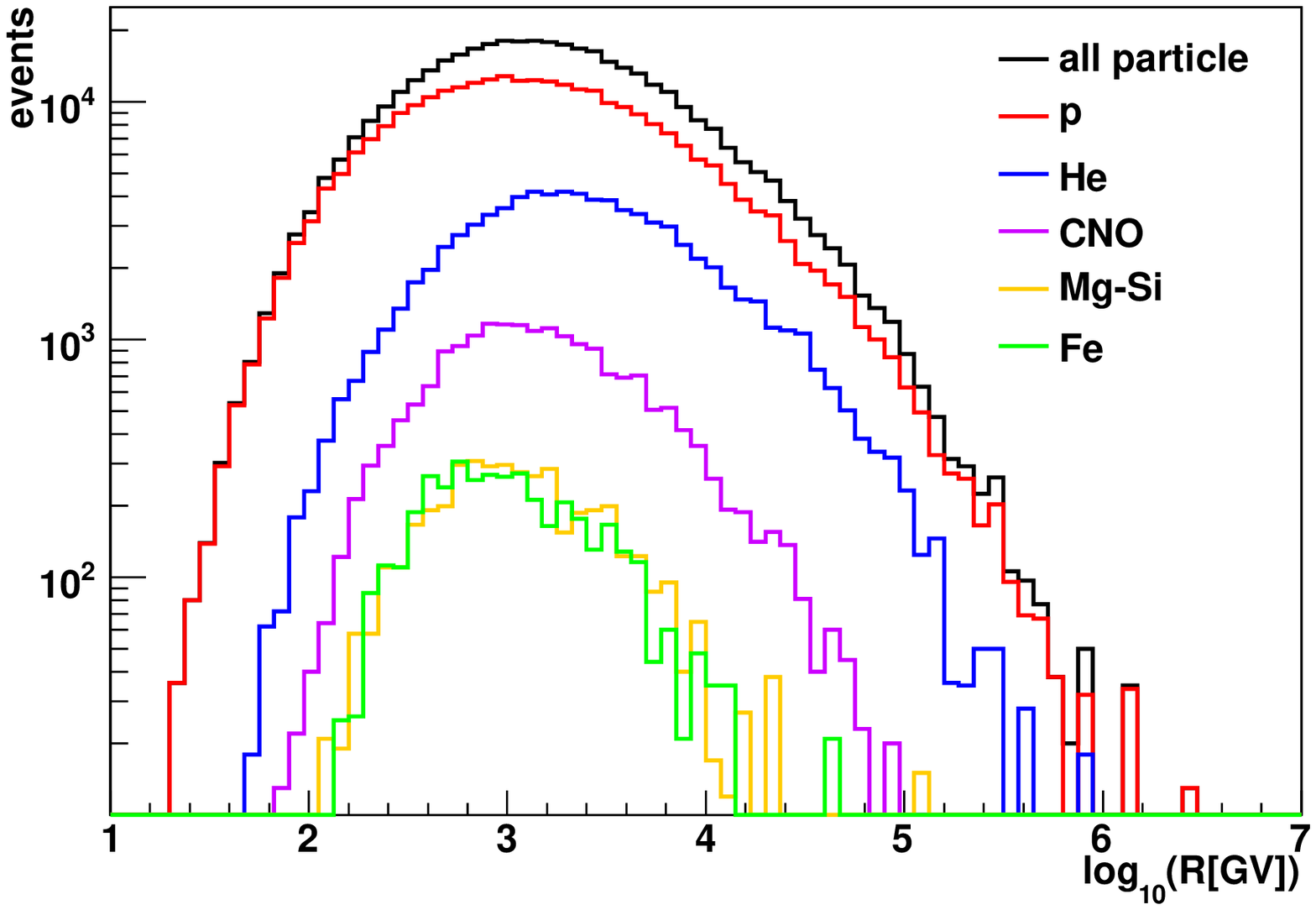}\\
  \caption{Rigidity distributions for events induced by different nuclei satisfying the selection criteria in Sec. II, B. The calculations refer to the showers of Fig. \ref{fig:components}.
\label{fig:enerigidity}}
\efi
%
\subsection{Role of the detector Point Spread Function}

The effects of the detector PSF and of the GMF can be studied separately in the East-West and North-South projections.
As already noticed, if we consider the magnetic deviation but not the smearing due to the angular resolution of the detector, the symmetry of the signal is broken only along the East-West direction (see Fig. \ref{fig:SimulatedMaps}, upper right map). 
Furthermore, the North-South deviation is less than Z$\cdot$ 0.1$^{\circ}$/E[TeV] for 95\% of CRs, making us confident that along this direction the signal is mostly affected by the angular resolution, which can be then determined. 

The angular width of the Moon (about half a degree) contributes to the spread of the signal, therefore we must disentangle this effect in measuring the detector angular resolution.
Assuming a Gaussian PSF with variance $\sigma_{\theta}^2$, the width of the observed signal results:
\begin{eqnarray}
\label{eq:rmsmoon} 
RMS = \sigma_{\theta}\sqrt{1+\left(\frac{r_{m}}{2\sigma_{\theta}}\right)^2} 
\end{eqnarray}
where $r_{m}$ is the Moon radius. The contribution of the Moon size to the RMS is dominant when $\sigma_{\theta}$ is low, i.e. at high particle multiplicities. For instance, the difference between $RMS$ and $\sigma_{\theta}$ is 
$20\%$ if $\sigma_{\theta}$ = 0.2$^{\circ}$, less than $5\%$ if $\sigma_{\theta} >$0.4$^{\circ}$, and only $1.7\%$ if $\sigma_{\theta}$ = 0.7$^{\circ}$.
%
\bfi{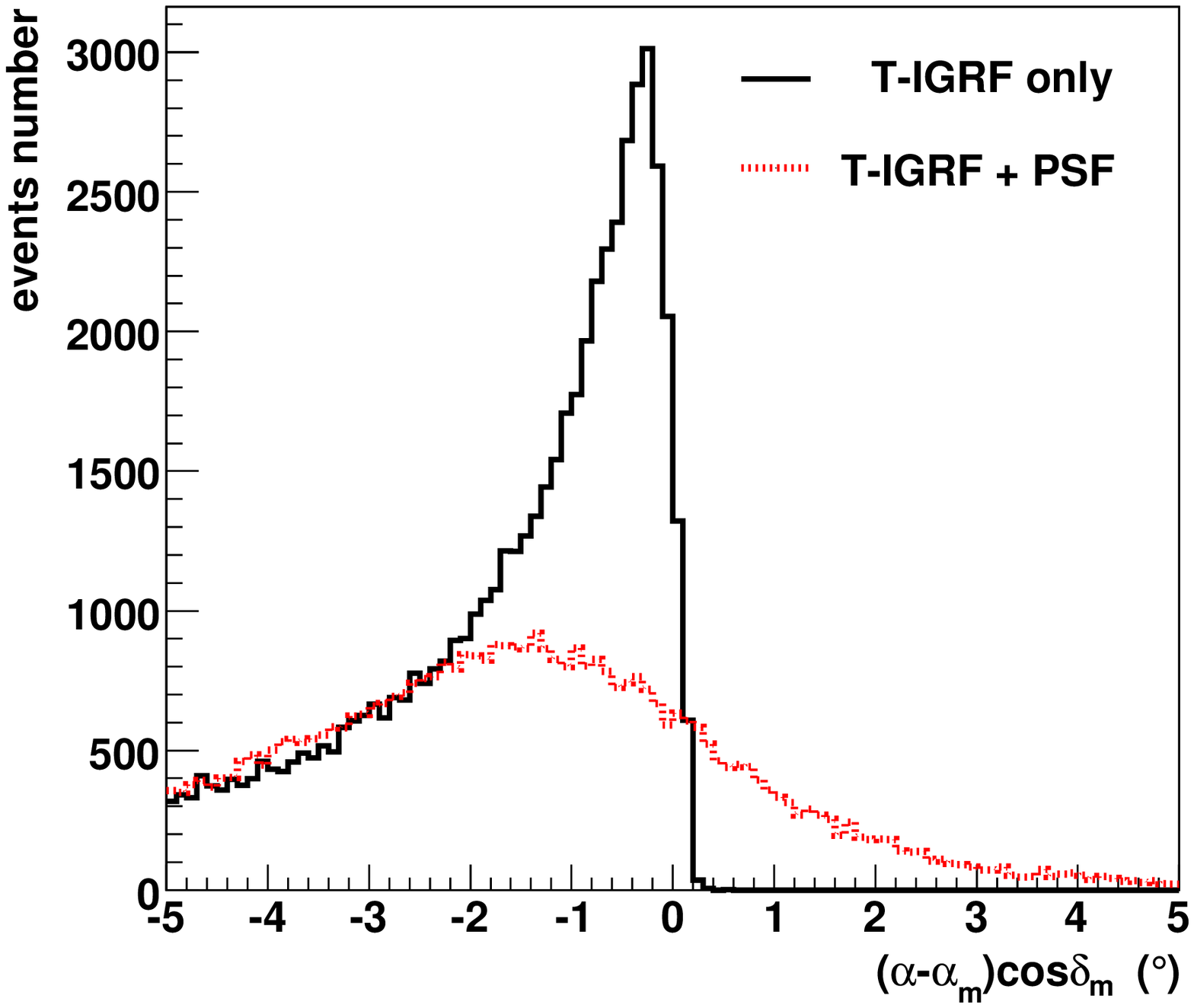}\\
  \caption{Effect of the detector PSF along the East-West direction for proton-induced showers. The continuous black line represents the Moon shadow deformed by the GMF as it would appear to an ideal detector. The segmented red line takes into account also the effect of the detector PSF: the diplacement of the signal peak results enhanced.
  \label{fig:EffectOfTheAngularResolution}}
\efi
%
In Fig. \ref{fig:EffectOfTheAngularResolution} the effect of the detector angular resolution along the East-West projection of the Moon shadow deficit is shown. Such an effect determines not only the smearing, but also a further displacement of the signal peak due to the folding with the asymmetrical deflection induced by the GMF. The West tail of the shifted signal, indeed, has a larger weight compared with the sharp East edge and tends to pull the signal in its direction.

\section{Results and discussions}

\subsection{The shape of the Moon shadow}

To get information on the detector performance the experimental shape of the Moon shadow for different shower multiplicities has been compared with the results of the Monte Carlo simulation of CR propagation in the Earth-Moon system.
The analysis is carried out by using the \emph{``source''} sky maps built subtracting the background maps to the event ones.
The deficit counts observed around the Moon projected on the
East-West axis are shown in Fig. \ref{fig:proj-ew} for 4 multiplicity bands compared to Monte Carlo expectations.
We used the events contained in an angular band parallel to the East-West axis and centered on the observed Moon position. The widths of these bands are chosen on the basis of the Monte Carlo simulation so that the shadow deficit is maximized. They turn out to be proportional to the N$_{strip}$-dependent angular resolution: $\pm$2.9$^{\circ}$ in
40$\leq N_{strip} <$ 60, $\pm$2.6$^{\circ}$ in 60$\leq N_{strip}
<$ 100, $\pm$2.1$^{\circ}$ in 100$\leq N_{strip} <$ 200,
$\pm$1.6$^{\circ}$ in 200$\leq N_{strip} <$ 500. 

As an expected effect of the GMF, the profile of the shadow is
broadened and the peak positions shifted westward as the
multiplicity (i.e., the cosmic ray primary energy) decreases. The
data are in good agreement with the expectations from the Monte Carlo
simulation discussed in the previous section. 

The deficit counts measured around the Moon projected along the North-South axis for the same multiplicity bins are compared to Monte Carlo expectations in Fig. \ref{fig:proj-ns}. As expected the shape of the deficit, not affected by the GMF along this direction, is symmetric and provides information about the detector PSF and the angular resolution. 
%
\bfi{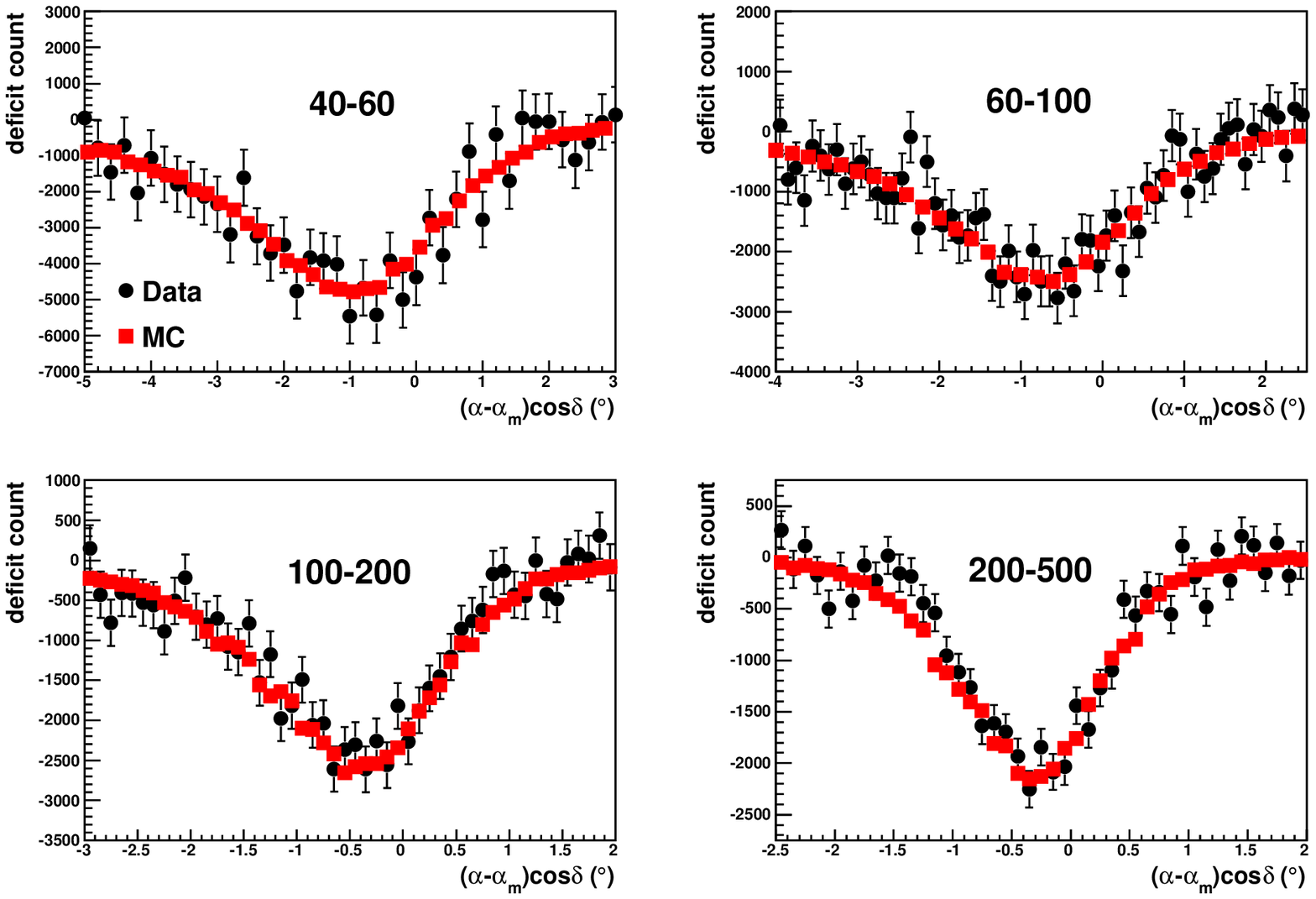}\\
  \caption{Deficit counts measured around the Moon projected along
the East-West axis for different multiplicity bins (black circles) compared to Monte Carlo expectations (red squares). Events contained in an angular band parallel to the East-West axis and centered on the observed Moon position, proportional to the multiplicity-dependent angular resolution, are used (see text).
  \label{fig:proj-ew}}
\efi
%
%
\bfi{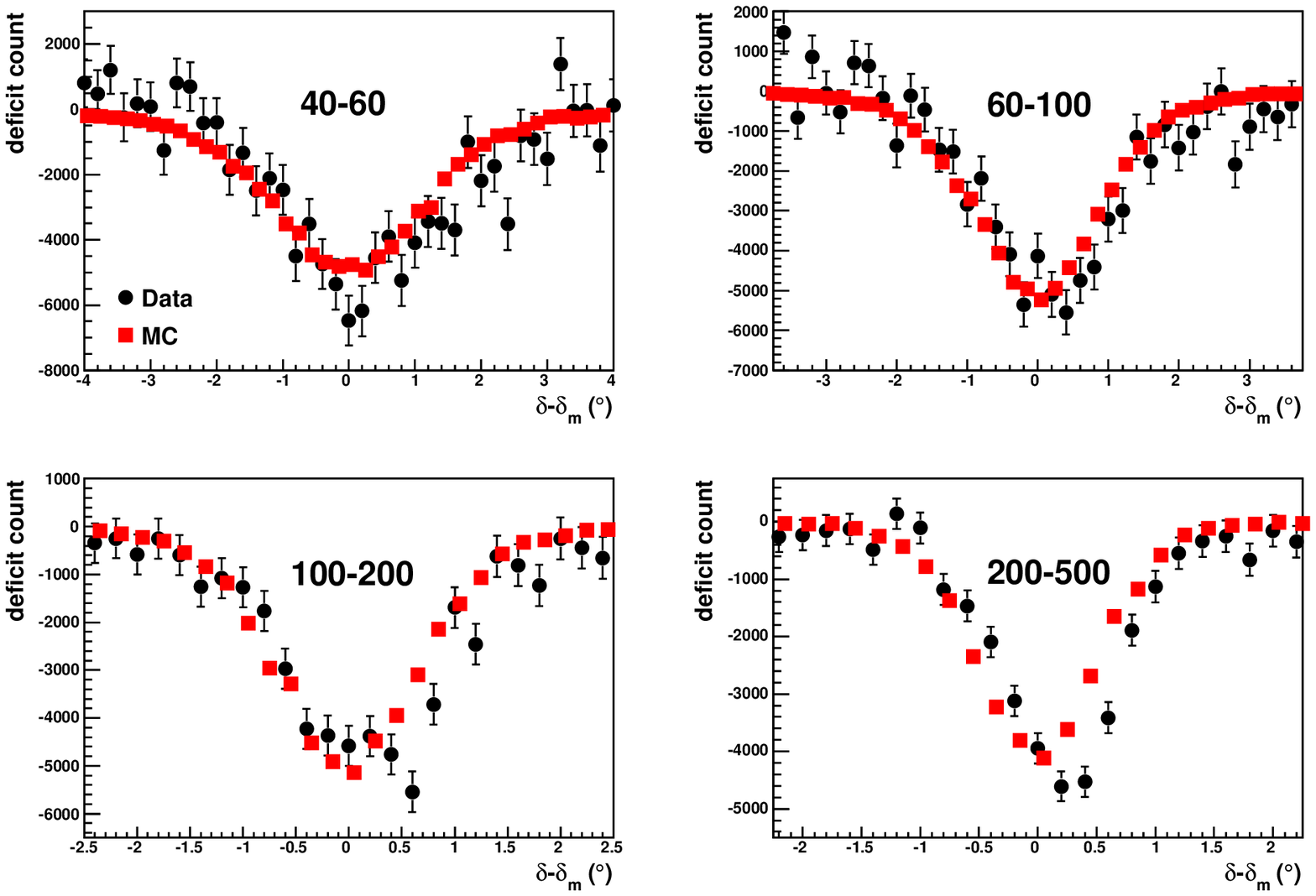}\\
  \caption{Deficit counts measured around the Moon projected along
the North-South axis for different multiplicity bins (black circles) compared to Monte Carlo expectations (red squares). Events contained in an angular band parallel to the North-South axis and centered on the observed Moon position, proportional to the multiplicity-dependent angular resolution, are used (see text).
  \label{fig:proj-ns}}
\efi
%
\bfi{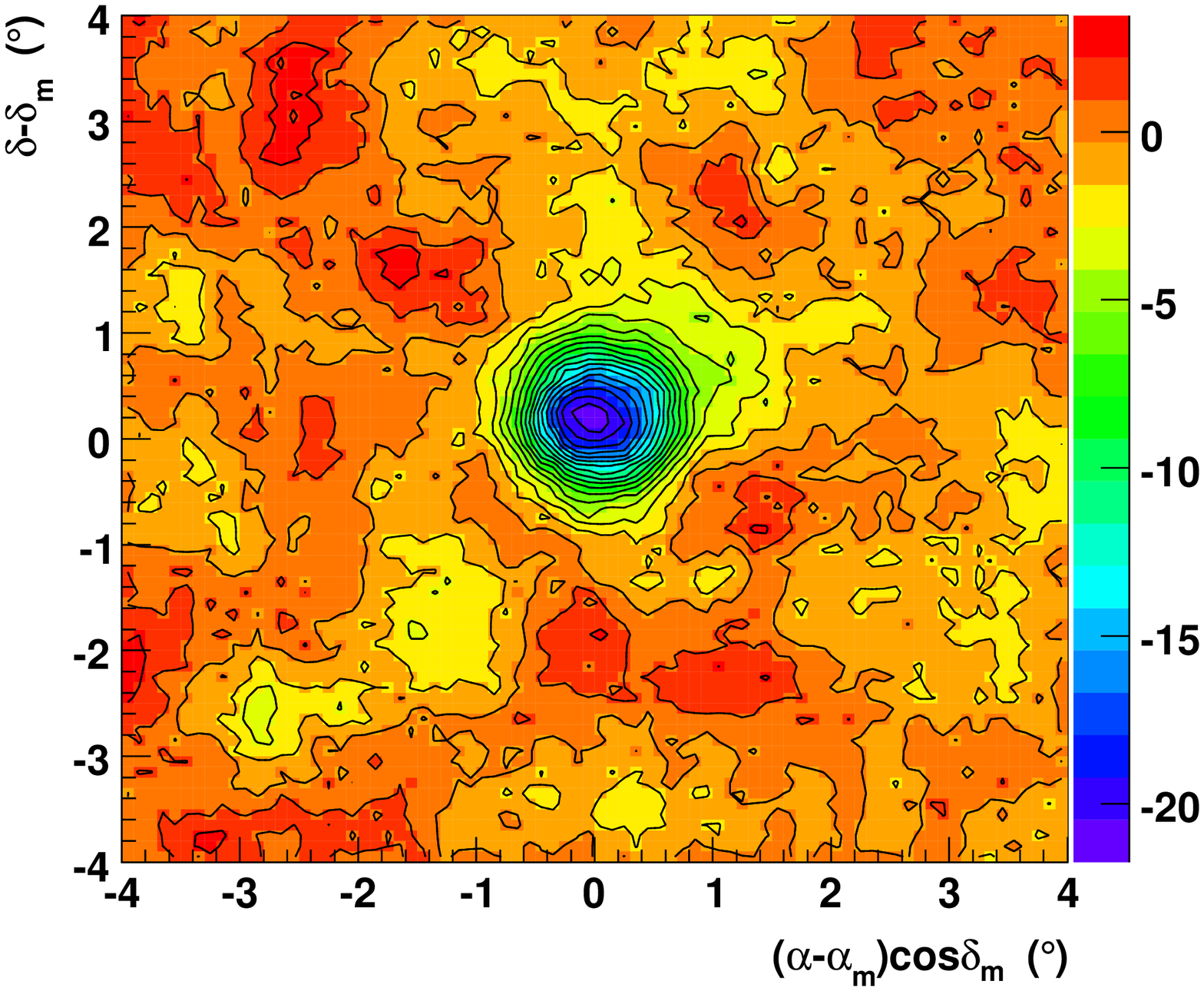}\\
  \caption{Significance map of the Moon shadow region observed by the ARGO-YBJ
detector in 3200 hours on-source for events with N$_{strip}\geq$1000. The color scale gives the statistical significance in terms of standard deviations.
  \label{fig:moon-hemap}}
\efi
%

\subsection{Pointing accuracy and angular resolution}

The pointing accuracy of the detector can be determined by observing the position of the Moon shadow cast by high energy CRs whose trajectory  is negligibly affected by the GMF.
The deflection of 30 TeV protons is about 0.05$^{\circ}$ (see Eq. (\ref{eq:DipoleDisplacement})). For heavier nuclei this deflection increases but as the composition of cosmic rays in this energy range is dominated by the light component (nuclei heavier than CNO contribute to the observed rate for less than 3\% in the whole strip multiplicity range \cite{catalanotti}), we expect only a negligible effect on the blurring of the Moon shadow from heavy ions.
The observed position of the Moon shadow cast by high energy (N$_{strip}\geq$1000, E$_{50}^p\sim$ 30 TeV) primary CRs is centered in the East-West direction, as shown in Fig. \ref{fig:moon-hemap}. 
On the other hand, according to the simulation of CR propagation in the Earth-Moon system, no displacement along the North-South direction at any energy is expected at the YangBaJing latitude. This analysis suggests that there is a residual systematic shift towards North independent of the multiplicity.
%
\bfi{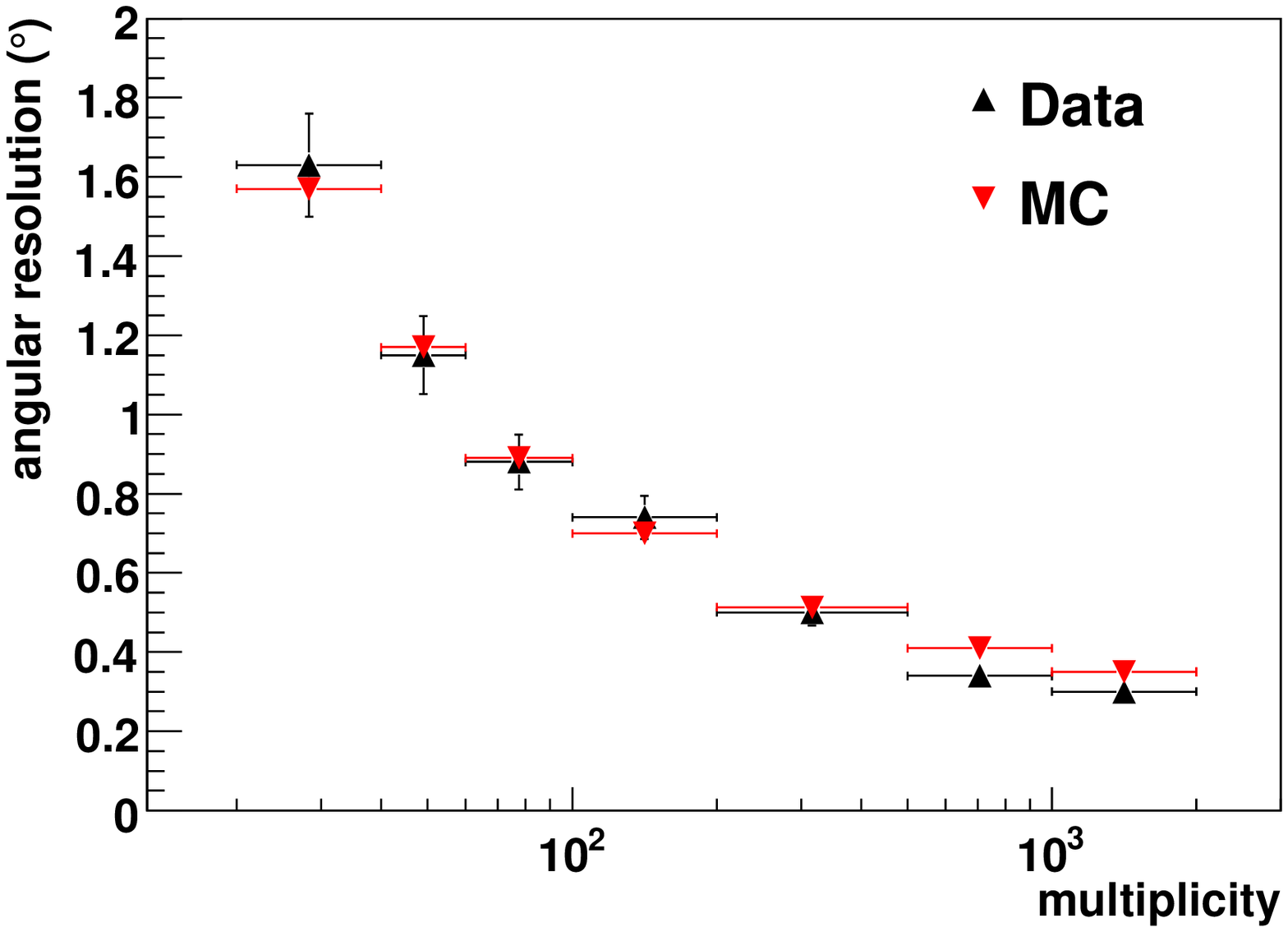}\\
  \caption{Measured angular resolution of the ARGO-YBJ detector (upward black triangles) compared to expectations from Monte Carlo simulation (downward red triangles) as a function of the particle multiplicity. The multiplicity bins are shown by the horizontal bars. 
\label{fig:angresol}}
\efi
%

The PSF of the detector, studied in the North-South projection not
affected by the GMF (see Sec. IV, C), is Gaussian for N$_{strip}\geq$200, while for lower multiplicities is better described for both Monte Carlo and data with a linear combination of two Gaussian functions. The second Gaussian contributes for about 20\%. For these events the angular resolution is calculated as the weighted sum of the $\sigma_{\theta}^2$ of each Gaussian.
In Fig. \ref{fig:angresol} the angular resolution measured along the North-South direction is compared to Monte Carlo predictions as a function of the particle multiplicity, i.e. the number of fired strips N$_{strip}$ on ARGO-130. The values are in fair agreement showing that the ARGO-YBJ experiment is able to reconstruct events starting from only 20 particles spread on an area $\sim$6000 m$^2$ large with an angular resolution better than 1.6$^{\circ}$.
The effect of the finite angular width of the Moon on the angular resolution, ruled by Eq. (\ref{eq:rmsmoon}), has been taken into account. 

This measured angular resolution refers to cosmic ray-induced air
showers. The same Monte Carlo simulation predicts an angular resolution for $\gamma$-induced showers smaller by $\sim$30-40$\%$, depending on  multiplicity, due to the better defined time profile of the showers. 

\subsection{Absolute rigidity scale calibration}

In order to calibrate the absolute rigidity scale of CRs observed by the ARGO-YBJ detector we can use the GMF as a magnetic spectrometer. 
In fact, the westward displacement of CRs by an angle inversely proportional to their energy (Eq. (\ref{eq:DipoleDisplacement})) provides a direct check of the relation between the shower size and the primary energy.
%
{\begin{figure}
\epsfxsize=9.5cm
\epsffile{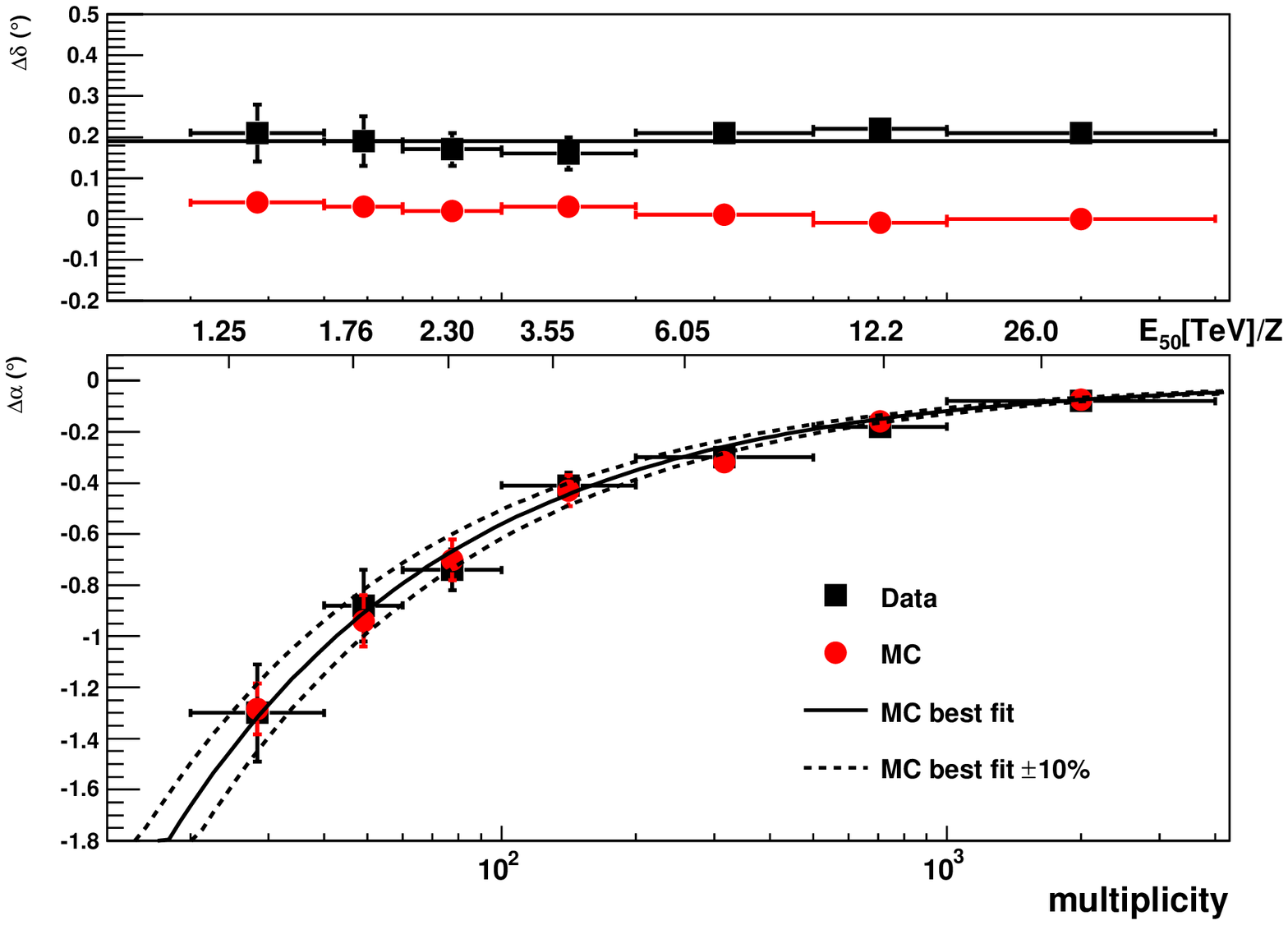}\\
\caption{Measured displacements of the Moon shadow (upper plot: North-South, lower plot: East-West) as a function of multiplicity (black squares). The data are compared to Monte Carlo expectations (red circles). In the upper plot the solid line is fitted to the data.
In the lower plot the solid curve is fitted to the Monte Carlo events and the dashed curves show the $\pm$10\% deviation from the solid one.
The rigidity scale refers to the rigidity (TeV/Z) associated to the median energy in each multiplicity bin (shown by the horizontal bars).
  \label{DataMCEW}}
\efi
%
In Fig. \ref{DataMCEW} the displacements of the Moon shadow in both North-South (upper plot) and East-West (lower plot) directions as a function of the particle multiplicity, i.e. the number of fired strips N$_{strip}$ on ARGO-130, are shown.
The rigidity scale refers to the rigidity (TeV/Z) associated to the median energy in each multiplicity bin.

The same Monte Carlo simulation predicts that at fixed multiplicity the median energy for $\gamma$-induced showers is smaller by $\approx$30$\%$ on average. 

The observed shift is compared to the results of the Monte Carlo simulation of CR propagation in the Earth-Moon system. 
A shift of (0.19$\pm$0.02)$^{\circ}$ towards North can be observed. This displacement is independent of the multiplicity.
Many tests on the absolute position of the detector, on the geometry of the experimental setup, on the time calibration and on the software for  reconstruction have been carried out. 
The most important contribution to the systematics is likely due to a residual effect not completely corrected by the time calibration procedure. Further studies are under way.

Concerning the East-West direction, the good agreement between data and simulation allows the attribution of this displacement to the combined effect of the detector PSF and the GMF. 
Therefore, the rigidity scale can be fixed in the multiplicity range 20-2000 particles, where the Moon shadow is moving under the bending effect of the GMF.
%
\bfi{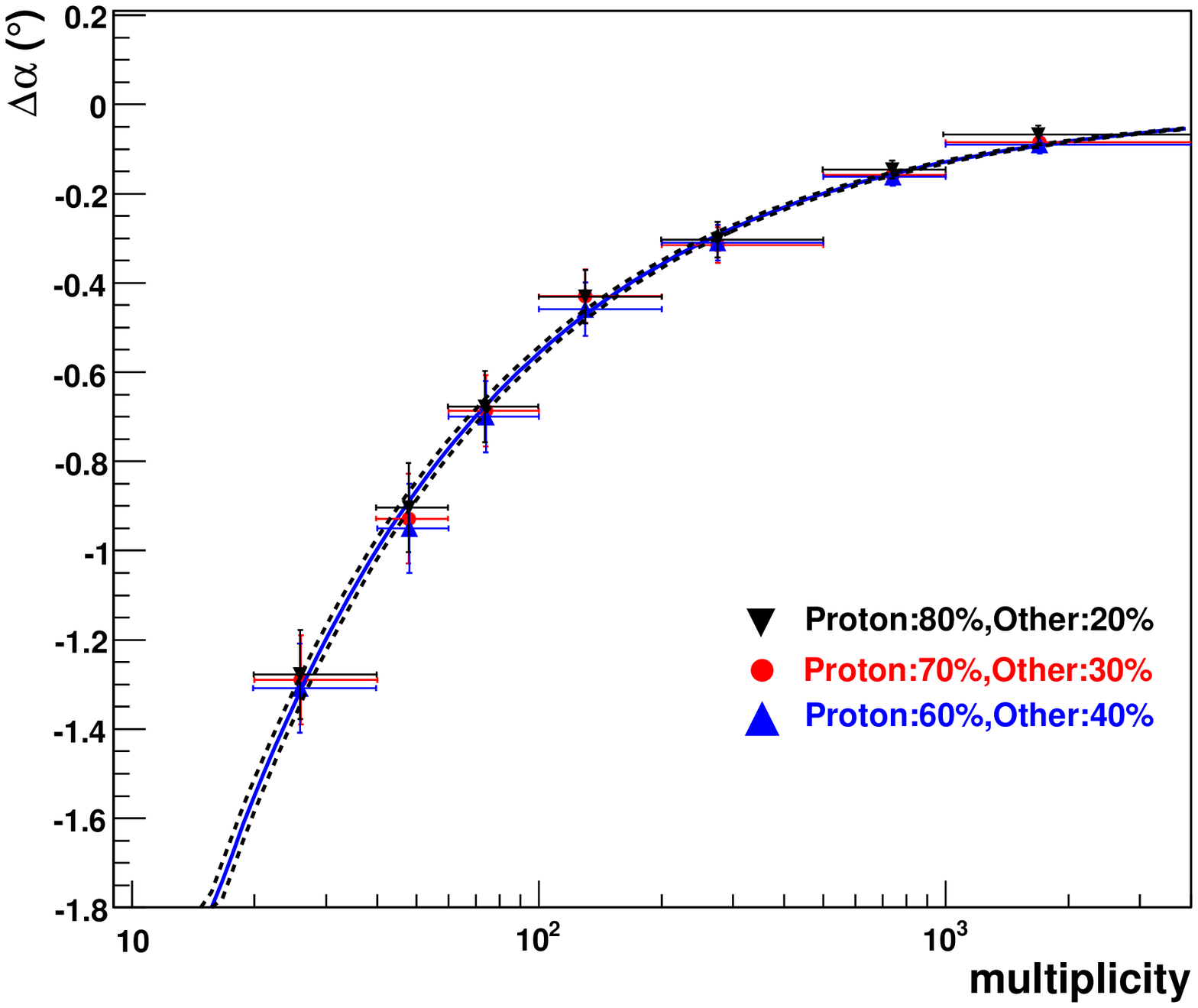}\\
    \caption{Expected westward displacement of the Moon shadow as a function of multiplicity, calculated assuming different primary composition models. The dashed curves show the 7$\%$ shift, corresponding to $\sigma_{chem}$, from the solid line (see text). The multiplicity bins are shown by the horizontal bars. 
  \label{DataMCpdiff}}
\efi
%
%
\bfi{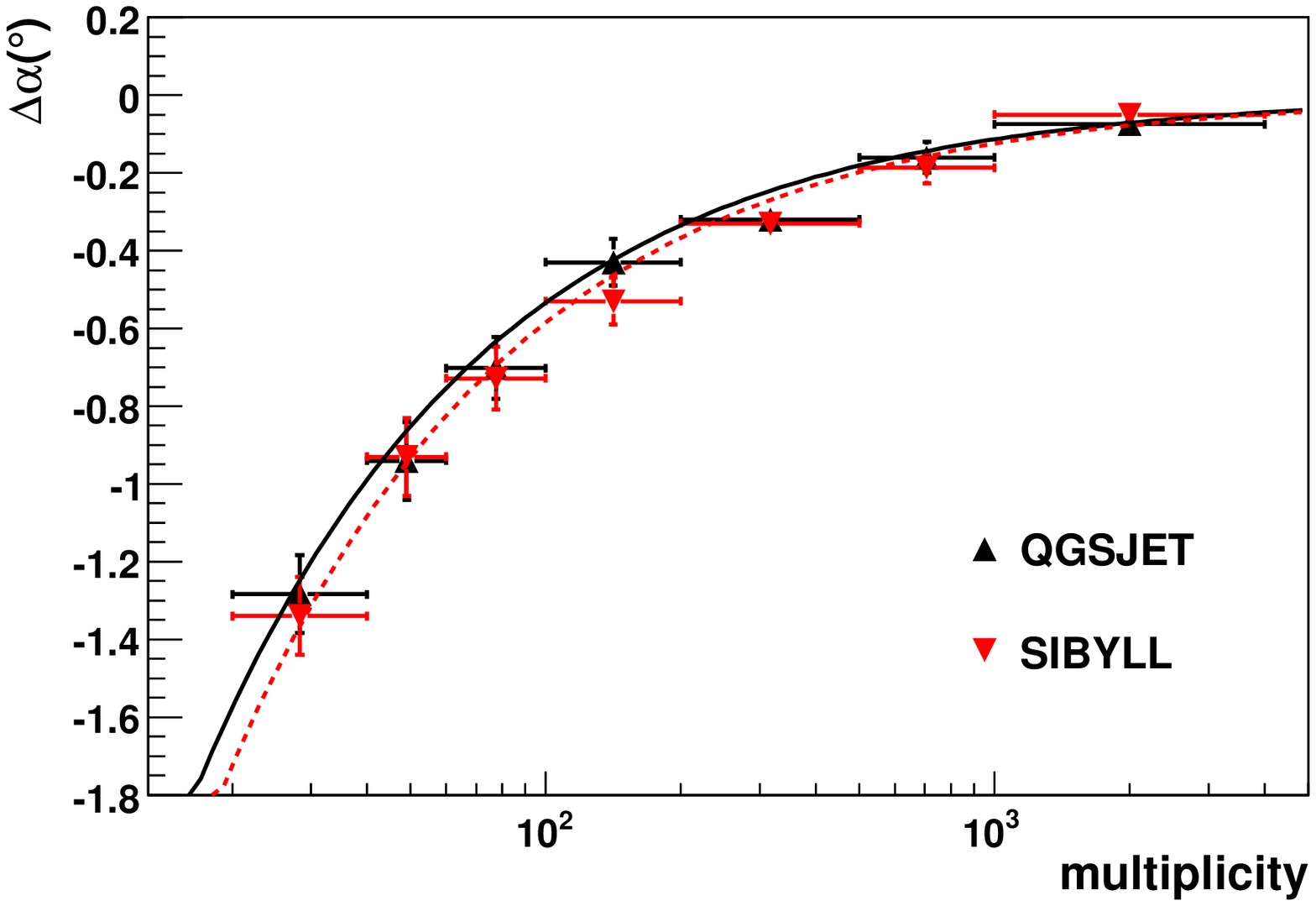}\\
  \caption{Expected westward displacement of the Moon shadow as a function of multiplicity calculated assuming different hadronic interaction models. The solid and dashed curves are the best-fit results assuming the QGSJET (upward black tringles) and SIBYLL (downward red triangles) models. The multiplicity bins are shown by the horizontal bars. 
  \label{DataMChaddiff}}
\efi
%
The Monte Carlo results are fitted by the function 
$\Delta\alpha=\kappa (N_{strip})^{\lambda}$, with $\kappa$ = -10.17 and $\lambda$ = -0.63, shown by the solid curve in Fig. \ref{DataMCEW}.
To estimate the possible shift in particle multiplicity between data and simulation, as shown by the dotted curves in Fig. \ref{DataMCEW}, the experimental data are fitted by the same function but with a multiplicity shift term:
    \begin{equation}
     -10.17[(1-\Delta R_{n}) N_{strip}]^{-0.63}
     \label{eq1}
    \end{equation}
as described in \cite{amenomori09}.
The parameter $\Delta R_{n}$ is the multiplicity shift ratio, resulting in $\Delta R_{n}$ = (+4 $\pm$ 7)$\%$.  
Finally, the conversion from $\Delta R_{n}$ to the energy shift ratio $\Delta R_{E}$ is performed. To determine the relationship between $\Delta R_{n}$ and
$\Delta R_{E}$, and to check that this method is sensitive to energy, six Monte Carlo event samples in which the energy of the primary particles is systematically shifted event by event in the Moon shadow simulation are calculated \cite{amenomori09}. These six $\Delta R_{E}$ samples correspond to $\pm 20\%$, $\pm 15\%$ and $\pm 8\%$. Finally, by assuming a linear dependence, the relation  $\Delta R_{n}$= (-0.91 $\pm$ 0.16)$\cdot\Delta R_{E}$ is obtained.
Hence, the systematic uncertainty in the absolute rigidity scale $\Delta R_{E}$ is estimated to be (+5 $\pm$ 8)$\%$, where the error is the statistical one.

Two systematic uncertainties may affect this analysis, the first related to the assumed primary CR composition.
In Fig. \ref{DataMCpdiff} the dependence of the Moon shadow displacement on the fraction of protons in the primary spectrum is shown as a function of the multiplicity. The proton ratio has been varied by $\pm$10$\%$ from the assumed standard chemical composition.
Indeed, in the investigated energy range, more than 90\% of the CRs triggering ARGO-YBJ are protons and He nuclei \cite{catalanotti}, whose spectra have been measured with uncertainties less than 10\% \cite{cream}.
The results have been fitted with function (\ref{eq1}) obtaining the systematic uncertainty associated to the chemical composition, $\sigma_{chem}$ = 7$\%$. 

The second source of systematic uncertainty may be related to the use of different hadronic interaction models. 
The results obtained with the QGSJet and SIBYLL codes are compared in 
Fig. \ref{DataMChaddiff}. 
The different displacements have been fitted with function (\ref{eq1}) obtaining the systematic uncertainty associated to these models, $\sigma_{hadr}$ = 12$\%$. We expect that this uncertainty will be reduced by using new hadronic codes developed on the basis of the LHC data.

Finally, the difference in the energy dependence of the Moon shadow displacement between data and Monte Carlo simulation has been estimated to be 
+5\% $\pm$ 8$_{stat}$\% $\pm$ 7$_{chem}$\% $\pm$ (12$_{hadr}$/2)\%.

The absolute rigidity scale uncertainty in the ARGO-YBJ experiment is estimated to be smaller than $\sqrt{\Delta R_{E}^{2} + \sigma^{2}_{stat} + \sigma^{2}_{chem} + {(\sigma_{hadr}/2)}^{2}}$ = 13$\%$ in the energy range from 1 to 30 (TeV/Z), where the Moon shadow is shifted from its position due to the effect of the GMF.

For $\gamma$-induced showers we expect a lower scale uncertainty due to lack of uncertainties related to the hadronic interaction models and to the primary composition.
%
\bfi{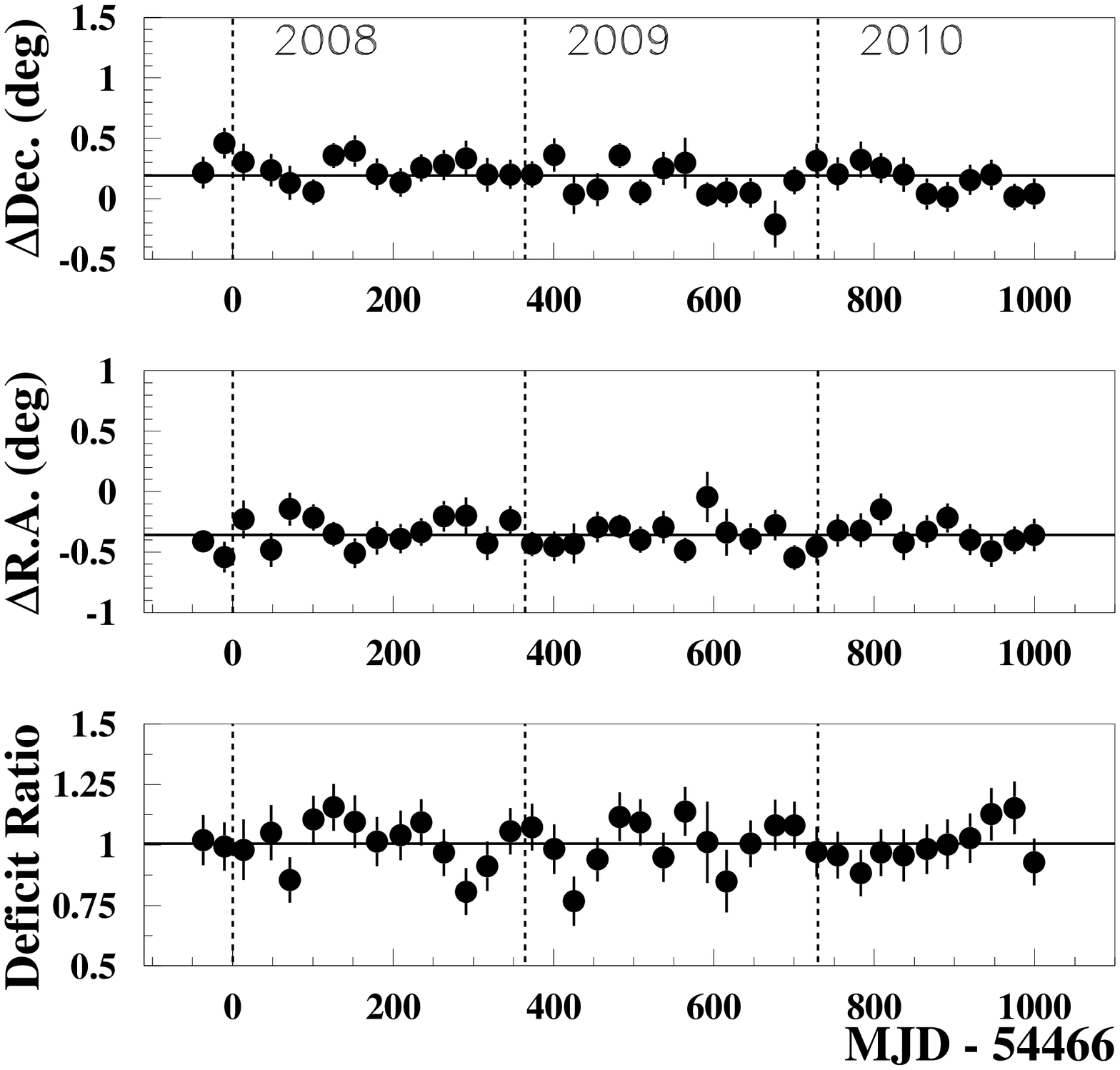}\\
  \caption{Upper panel: Displacement of the Moon shadow from the apparent center in the North-South direction as a function of the observation time.
  Middle panel: Displacement of the Moon shadow from the apparent center in the East-West direction as a function of the observation time.
  Lower panel: The ratio of the observed deficit count to the expected one as a function of the observation time. The plots refer to events with a multiplicity N$_{strip}>$100.
  \label{Data-stability}}
\efi
%
\bfi{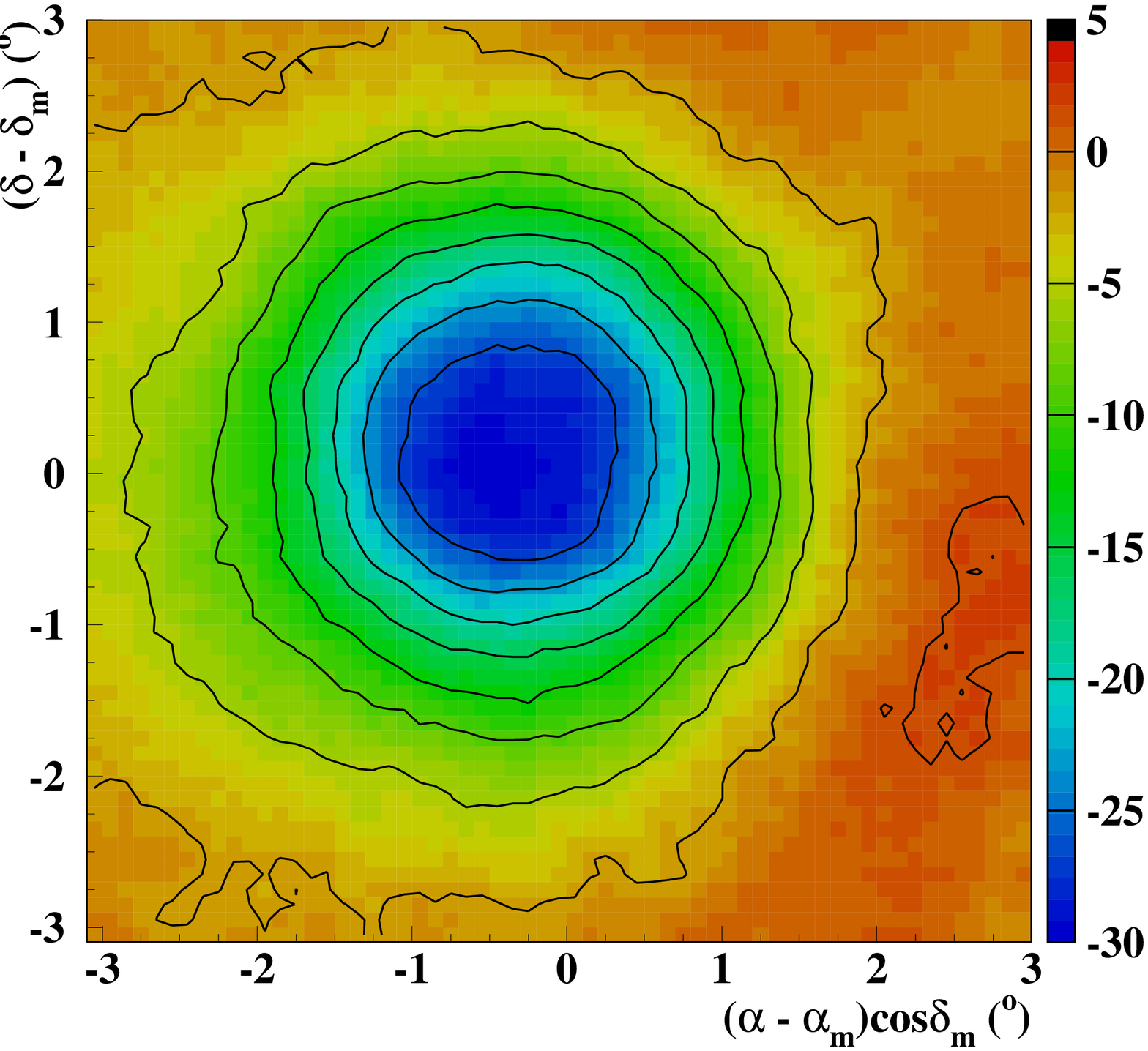}\\
(a)\\
\epsfxsize=9cm \epsffile{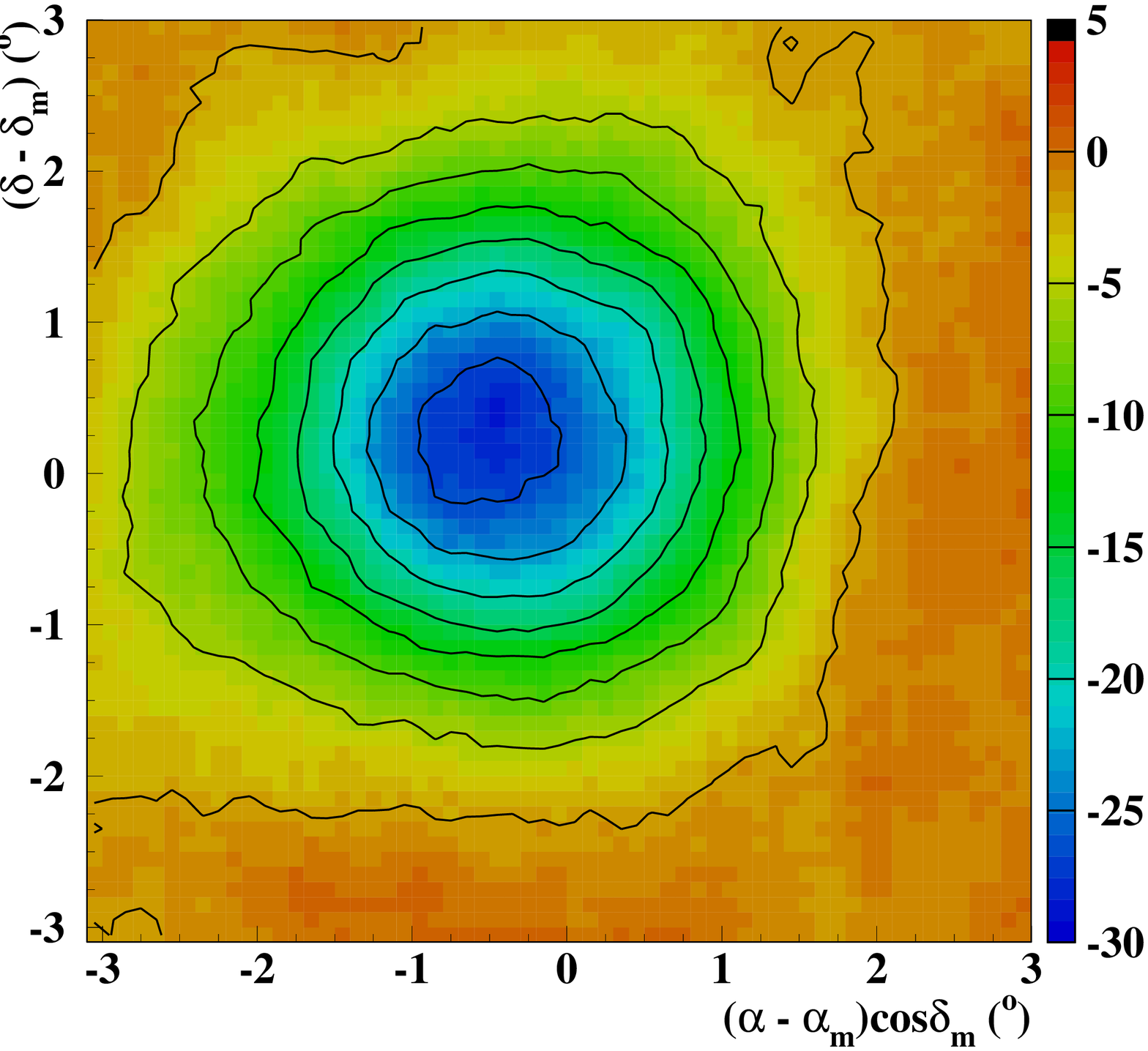}\\
(b)\caption{Significance map of the Moon region observed during "day" (panel (a)) and "night" (panel (b)) for events with N$_{strip}>$60. The color scale gives the statistical significance in terms of standard deviations.
  \label{fig:day-night}}
\efi
%

\subsection{Long-term stability of the detector}

The stability of the detector performance as far as the pointing accuracy and 
the angular resolution are concerned is crucial in $\gamma$-ray astronomy.
Since November 2007 the full detector is in stable data taking with duty cycle $\geq$85\%.
Therefore, the stability of the ARGO-YBJ experiment has been checked by monitoring both the position of the Moon shadow, separately along R.A. and DEC. projections, and the amount of shadow deficit events in the period November 2007 - November 2010, for each sidereal month and for events with N$_{strip}>$100. 

As discussed in Sec. V, B, the displacement of the center of the Moon shadow in the North-South direction enables us to estimate the systematic error in pointing accuracy and its long-term stability aside from Monte Carlo simulations, since the East-West component of the GMF is almost zero at YangBaJing. The displacement of the shadow position from the Moon center in the North-South direction is plotted in the upper panel of Fig. \ref{Data-stability} as a function of the observation time. 
Assuming a constant function, the best-fit result (continuous line) shows that the Moon shadow is shifted towards North by (0.19$\pm$0.02)$^{\circ}$. The RMS around this position is 0.13$^{\circ}$.

In the middle plot the displacement along the East-West direction is shown. The best-fit result (continuous line) shows that the Moon shadow is shifted towards West by (-0.36$\pm$0.02)$^{\circ}$, in agreement with the Monte Carlo expectations ((-0.35$\pm$0.07)$^{\circ}$). The RMS around this Moon position is 0.11$^{\circ}$.

The amount of CR deficit due to the Moon provides a good estimation of the size of the shadow, therefore of the angular resolution. 

The observed number of deficit events N$_{def}(<R)$ within an angular distance R from the Moon center is approximately given by
    \begin{equation}
N_{def}(<R)=\left[1-e^{-0.5\cdot(\frac{R}{\sigma})^2}\right]\cdot N_{moon}
\label{eq:ratio}
    \end{equation}
where N$_{moon}$ is the number of events intercepted by the Moon and $\sigma$ is the Gaussian width of the shadow. N$_{moon}$ is estimated by fitting the Eq. \ref{eq:ratio} to the experimental points fixing $\sigma$=0.8$^{\circ}$. The expected deficit events are simply counted from the background content within the Moon disc.
In the lower plot of Fig. \ref{Data-stability} the ratio of the observed 
deficit count to the expected one is shown as a function of the observation time.
The line shows the best-fit result assuming a constant function: 1.005$\pm$0.016. The RMS of the corresponding distribution is 0.11.

In conclusion, the position of the Moon shadow measured with the ARGO-YBJ experiment turned out to be stable at level of 0.1$^{\circ}$ and the angular resolution is stable at level of 10\%, on a monthly basis.

\subsection{The day/night effect}

Some differences in the Moon shadow signal could arise due to the interaction
of the GMF with the solar wind. We can point them out by dividing the data sample according to the relative positions of the Moon-Sun system, 
i.e. by requiring that the angular distance between the Moon and the Sun is 
smaller or larger than 90$^{\circ}$. This selection can be carried out by dividing all recorded data in two "day"/"night" samples. 

A similar analysis has been performed by the Tibet AS$\gamma$ array using 1990 - 1992 data \cite{tibet-daynight} and by the MACRO underground detector with high-energy muons in the period from 1989 to 2000 \cite{macro03}, though at higher energies (Tibet AS$\gamma$ at $\sim$10 TeV and MACRO at $\sim$20 TeV) and with less statistics.
While the Tibet collaboration did not find any day-night effect, the MACRO experiment observed a sharper shadow in the "night" sample and a broader shadow in the "day" one, with a lower significance, concluding that the night events encounter a reduced magnetic field with respect to the day events.

The results of the analysis carried out with the ARGO-YBJ experiment for the two subsamples are shown in Fig. \ref{fig:day-night} for events with N$_{strip}>$60. The data refer to the November 2007 -- November 2009 period and in each sample the Moon is observed for about 1150 hours.
We did not find any appreciable difference between the ``day'' and the ``night'' shadows (statistical significance of the maximum deficit 31 vs 30 s.d.) shadows. Accordingly, the shape of the Moon shadow seems to be independent of the position of the Moon with respect to the Sun. This implies that effects due to the solar wind do not give a considerable contribution to the CR bending, at least in the period of minimum of the solar activity.

\section{Conclusions}

The Moon shadowing effect on cosmic rays has been observed by
the ARGO-YBJ experiment in the multi-TeV energy region with a statistical significance greater than 55 standard deviations.

We observed a westward displacement, due to the GMF, up to about 1.3$^{\circ}$, proving the detection of the Moon shadow cast also by sub-TeV primaries. By means of an accurate Monte Carlo simulation of the CR propagation in the Earth-Moon system we have studied in detail the role of the GMF and of the detector PSF on the observed shadow.
The measured deficit counts around the Moon position are found in fair agreement with the expectations based on the primary cosmic ray composition derived from the direct observational data. 

The dependence of the measured angular resolution on the particle multiplicity is in good agreement with Monte Carlo calculations. A systematic shift of (0.19$\pm$0.02)$^{\circ}$ towards North has been observed.

We have estimated the primary energy of the detected showers by measuring the westward displacement as a function of the multiplicity, thus calibrating the relation between shower size and CR energy.
The systematic uncertainty in the absolute rigidity scale is evaluated to be less than 13\% in the range from 1 to 30 (TeV/Z), mainly due to the statistical one. 

The position of the Moon shadow measured with the ARGO-YBJ experiment turned out to be stable at a level of 0.1$^{\circ}$ and the angular resolution stable at a level of 10\%, on a monthly basis.
These results make us confident about the detector stability in the long-term observation of gamma-ray sources.

Finally, we have studied with high statistical accuracy the shadowing effect in the "day"/"night" time looking for possible effects induced by the solar wind. Within the statistical accuracy of this study we find that the solar wind does not give appreciable contribution to the CR bending, at least in the period of minimum of the solar activity.

\begin{acknowledgments}
This work is supported in China by NSFC (No. 10120130794), the Chinese Ministry of Science and Technology,
the Chinese Academy of Sciences, the Key Laboratory
of Particle Astrophysics, CAS, and in Italy by the Istituto
Nazionale di Fisica Nucleare (INFN). We also acknowledge
the essential supports of W. Y. Chen, G. Yang, X. F. Yuan, C.
Y. Zhao, R. Assiro, B. Biondo, S. Bricola, F. Budano, A. Corvaglia,
B. D'Aquino, R. Esposito, A. Innocente, A. Mangano,
E. Pastori, C. Pinto, E. Reali, F. Taurino, and A. Zerbini, in the
installation, debugging, and maintenance of the detector.
\end{acknowledgments}

\section*{Appendix}

In this Appendix the analytical calculation of Eq.
(\ref{eq:DipoleDisplacement}) is presented.
Since only the magnetic field is supposed to act upon the particles trajectories, they read as:
\begin{equation}
 \label{eq:IntegralLorentzEquation}
\mathbf x(t)=\mathbf x_0+\mathbf
v_0\,t+\frac{Zec^2}{E}\int_0^t\textrm{d}\tau\,\int_0^\tau\textrm{d}\alpha\,
\frac{\textrm{d}\mathbf x}{\textrm{d} \alpha}\times \mathbf
B(\mathbf x,\alpha)
\end{equation}
at time $t$ in a certain reference frame, where:
\begin{itemize}
 \item $\mathbf x(t)$ is the particle position at time $t$;
 \item $\mathbf x_0$ and $\mathbf v_0$ are the initial position and velocity of
the particle;
 \item $Ze$ and $E$ are its charge and its (constant) energy;
 \item $\mathbf B(\mathbf x,t)$ is the magnetic field, which in principle can
vary with respect to both position and time;
\item $\alpha$ is the time the inner integral is computed over.
\end{itemize}

If it is possible to write an explicit functional form for $B(\mathbf x,t)$, an attempt to solve Eq. (\ref{eq:IntegralLorentzEquation}) can be made. On the contrary, especially when no analytical expression of the time behavior is known, the equation can be solved with numerical techniques. 

Equation (\ref{eq:IntegralLorentzEquation}) explicitly shows the
perturbation induced by the magnetic field on the straight
trajectory ($\mathbf x(t) = \mathbf x_0+\mathbf v_0t$). It
suggests an iterative method to determine the solution, which can
be expressed as the series:
\begin{eqnarray}
 \label{eq:SolutionSeries}
\mathbf x(t)= \mathbf x_{\mathcal O(B^0)}(t)+\mathbf x_{\mathcal
O(B^1)}(t)+\dots\nonumber
\end{eqnarray}

where $x_{\mathcal O(B^0)}(t)=\mathbf x_0 +\mathbf v_0t$ is the
unperturbed (straight) trajectory and for higher orders we have:

\begin{eqnarray*}
\label{eq:SolutionSeriesTerm}
\Delta\mathbf x_{\mathcal O(B^{i+1})}(t)=\\
=\frac{Zec^2}{E}\int_0^t\textrm{d}\tau\int_0^\tau\textrm{d}\alpha
\frac{\textrm{d}\mathbf x_{\mathcal O(B^i)}}{\textrm{d}
\alpha}\times \mathbf B(\mathbf x_{\mathcal O(B^i)},\alpha)\\
\textrm{\ \ \ $i=0,1,\dots$} 
\end{eqnarray*}

where $\Delta\mathbf x_{\mathcal O(B^{i+1})}(t)=\mathbf x_{\mathcal O(B^{i+1})}(t)-(\mathbf x_0+\mathbf v_0\,t)$
is the displacement from the unperturbed trajectory at time $t$. At the first-order approximation we find:
\begin{eqnarray*}
\Delta\mathbf x(t)\simeq\frac{Zec^2}{E}\mathbf
v_0\times \int_0^t\textrm{d}\tau\,\int_0^\tau\textrm{d}\alpha\,
\mathbf B(\mathbf x_0+\mathbf v_0\alpha,\alpha)
\end{eqnarray*}
or:
\begin{eqnarray}
 \label{eq:FirstOrderDisplacement}
\Delta\mathbf x(t)\simeq\frac{Z}{E}\mathbf
v_0\times\mathcal I_{\mathbf B}(t;\mathbf x_0, \mathbf v_0)
\end{eqnarray}
where $\mathcal I_{\mathbf B}(t;\mathbf x_0, \mathbf v_0)$ is the integral of the magnetic field along the straight trajectory, whose value depends only on the time of the motion ($t$) and on its initial conditions ($\mathbf x_0$ and $\mathbf v_0$).

Since the phenomenon studied concerns ultrarelativistic particles, once we fix the initial position and the final time, Eq. (\ref{eq:FirstOrderDisplacement}) reads:
\begin{eqnarray}
\Delta\mathbf x\simeq\frac{Z}{E}\ \hat{\mathbf v}_0\times\mathcal
I_{\mathbf B}(\hat{\mathbf v}_0)\nonumber
\end{eqnarray}
At the first approximation the displacement depends only on the charge-to-energy ratio of the primary and on the initial direction of its ultrarelativistic motion (versor $\hat{\mathbf v}_0$).

Now, let us consider only the lowest order of the GMF multipoles-expansion, i.e. the \emph{dipole term}:
\begin{eqnarray}
\mathbf B(\mathbf x)=\frac{3(\mathbf b\cdot\mathbf x)\mathbf
x-x^2\mathbf b}{x^5}\nonumber
\end{eqnarray}
where $\mathbf b$ has intensity $b\approx8.1\cdot10^{27}$ T m$^3$ 
and the south magnetic pole is supposed to have coordinates 
$78.3^{\circ}$ South, $111.0^{\circ}$ East.
By setting $\hat{\mathbf v}_0||\mathbf x_0$ (\emph{vertical direction approximation}) and integrating from YangBaJing to a distance $\sim60$ Earth radii, Eq. (\ref{eq:DipoleDisplacement}) is immediately obtained.


\end{document}